\documentclass[12pt]{spieman}  
\usepackage{amsmath,amsfonts,amssymb}
\usepackage{graphicx}
\usepackage{setspace}
\usepackage{tocloft}
\usepackage{filecontents}
\usepackage{siunitx}
\usepackage{epstopdf}
\usepackage{booktabs}
\usepackage{multirow}
\usepackage{float}
\usepackage{subfig}
\usepackage{enumitem}
\usepackage[nomarkers]{endfloat}

\title{Demonstration of broadband contrast at 1.2 $\lambda$/D and greater for the EXCEDE Starlight Suppression System}

\author[a]{Dan Sirbu}
\author[b]{Sandrine J. Thomas}
\author[a]{Ruslan Belikov}
\author[c]{Julien Lozi}
\author[a]{Eduardo Bendek}
\author[a]{Eugene Pluzhnik}
\author[a]{Dana H. Lynch}
\author[d]{Troy Hix}
\author[a]{Peter Zell}
\author[c,e]{Olivier Guyon}
\author[e]{Glenn Schneider}
\affil[a]{NASA Ames Research Center, Moffett Field, CA, 94035} 
\affil[b]{Large Synoptic Survey Telescope, Tucson, AZ, 85719}
\affil[c]{National Astronomical Observatory of Japan, Subaru Telescope, Hilo, HI, 96720}
\affil[d]{Lockheed Martin Space Systems Company, Palo Alto, CA, 94304}
\affil[e]{The University of Arizona, Tucson, AZ, 85721}

\cftpagenumbersoff{figure}
\cftpagenumbersoff{table} 
\begin{document} 
\maketitle

\begin{abstract}
The EXoplanetary Circumstellar Environments and Disk Explorer (EXCEDE) science mission concept uses a visible-wavelength Phase-Induced Amplitude Apodization (PIAA) coronagraph to enable high-contrast imaging of circumstellar debris systems and some giant planets at angular separations reaching into the habitable zones of some of the nearest stars. We report on the experimental results obtained in the vacuum chamber at the Lockheed Martin Advanced Technology Center in 10\% broadband light centered about 650 nm, with a median contrast of $1 \times 10^{-5}$ between 1.2 and 2.0 $\lambda$/D simultaneously with $3 \times 10^{-7}$ contrast between 2 and 11 $\lambda/D$ for a single-sided dark hole using a deformable mirror (DM) upstream of the PIAA coronagraph. These results are stable and repeatable as demonstrated by three measurements runs with DM settings set from scratch and maintained on the best 90\% out of the 1000 collected frames. We compare the reduced experimental data with simulation results from modeling observed experimental limits; observed performance is consistent with uncorrected low-order modes not estimated by the Low Order Wavefront Sensor (LOWFS). Modeled sensitivity to bandwidth and residual tip/tilt modes is well-matched to the experiment. 
\end{abstract}


{\noindent \footnotesize\textbf{Address all correspondence to:}\\ Dan Sirbu, NASA Ames Research Center, Moffett Field, Mountain View, CA, 94035;  \linkable{dan.sirbu@nasa.gov} \\ Glenn Schneider, The University of Arizona, Tucson, AZ, 85721; \linkable{gschneid@email.arizona.edu}}

\begin{spacing}{2}   

\section{Introduction}
\label{sect:intro}

The EXoplanetary Circumstellar Environments and Disk Explorer (EXCEDE) is a proposed Explorer mission concept \cite{guyon12}. EXCEDE was selected by NASA as a Category III laboratory investigation for an experimental demonstration of the underlying high-contrast imaging technology fulfilling a key goal in the Astro2010 Decadal Review to mature the technology capability for direct imaging of circumstellar debris systems and exoplanets. \cite{astro2010}.

The EXCEDE mission is a science-driven technology-pathfinder. It uses a 0.7 \si{\meter} diameter off-axis three-mirror anastigmat telescope with an unobscured pupil, that images in two $\leq$20\%-wide bands at 0.4 and 0.8 \si{\micro \meter}. The main technological challenge of the coronagraph instrument is to achieve a high-level of contrast at a small angular separation (also called stellocentric angle) from the host star (starting at a separation of 1.4 AU from a star at 10 pc). The EXCEDE starlight suppression system (SSS) has a science-driven $10^{-7}$ raw contrast requirement for an angular separation between 2 and 22 $\lambda$/D and $10^{-6}$ raw contrast from 1.2 to 2 $\lambda$/D. The EXCEDE back-end science camera is a (two-band) Nyquist-sampled full-linear Stokes imaging polarimeter that (in addition to providing compositional diagnostics and geometrical constraints on circumstellar material) provides order-of magnitude improvement over raw contrast with polarized intensity imaging \cite{Schneider14}, independent of additional post-processing improvements by speckle calibration and other methods.These capabilities enable EXCEDE's primary science mission goals to directly image and characterize the circumstellar light-scattering material and (if present and sufficiently bright) giant exoplanets beyond and into stellar habitable zones. 


\begin{figure}[t!]
\centering
\includegraphics[height = 2.0 in]{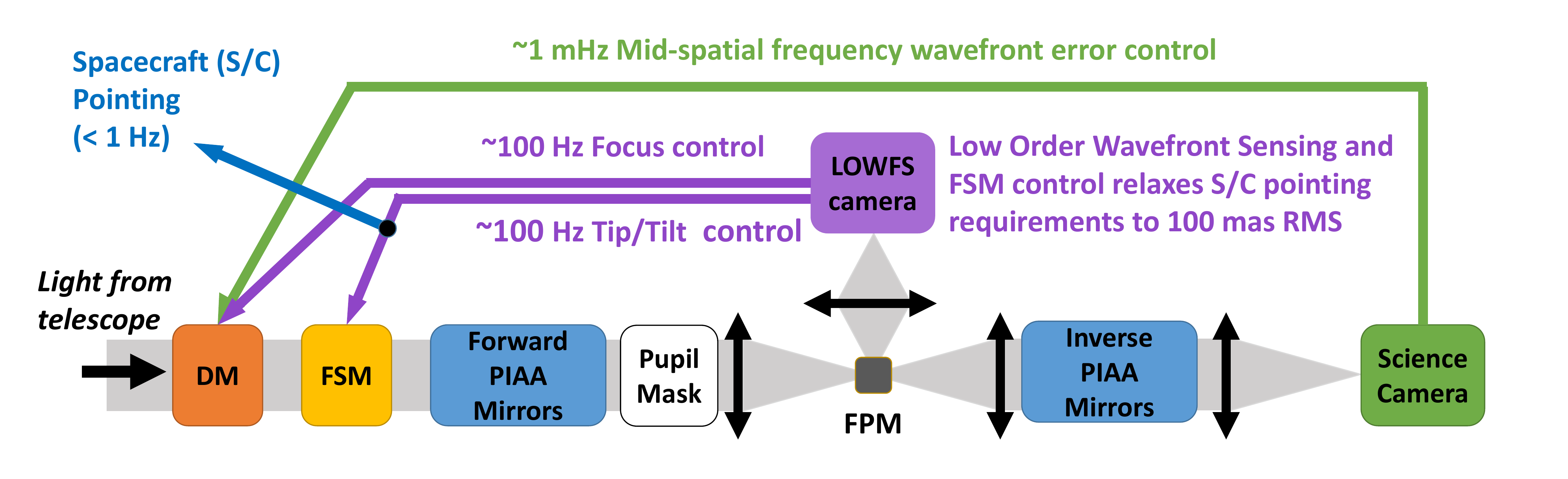}
\caption[EXCEDE flight-concept starlight suppression system block diagram.]{\label{fig:excedeSSS} Block diagram of the EXCEDE flight-concept Starlight Suppression System. A Low Order Wavefront Sensor (LOWFS) camera is used in conjunction with a Deformable Mirror (DM) and a Fast Steering Mirror (FSM) to remove low-order modes upstream from the Phase-Induced Amplitude Apodization (PIAA) coronagraph. The PIAA coronagraph consists of forward PIAA mirrors with a Focal Plane Mask (FPM). Inverse PIAA mirrors enable a wide field of view. The mid-spatial frequency wavefront correction loop using estimates from the science camera generates a dark hole in the image plane enabling high-contrast imaging.}
\end{figure}

The proposed flight-concept EXCEDE SSS is shown in block-diagram form in Figure \ref{fig:excedeSSS} and addresses the two fundamental physical phenomena that prevent direct imaging of circumstellar debris systems and exoplanets with an off-axis, non-segmented space telescope: diffraction effects due to the finite aperture and speckle-generation due to optical aberrations.

First, diffraction of the central starlight by an unobscured circular aperture creates the well-known Airy disk point spread function (PSF) whose innermost rings are several orders of magnitude brighter than the science targets which are then lost in the glare of the PSF halo. A key component of the SSS is a set of two specially-shaped mirrors that act as a Phase-Induced Amplitude Apodization (PIAA) coronagraph. The surface curvature of these PIAA mirrors reshape the input light beam to a more favourable apodization that reduces the amplitudes of the diffraction rings of the PSF \cite{guyon05}. A focal plane occulter then blocks the central core of the apodized PSF providing starlight suppression and further improving contrast in the (later re-imaged) focal-plane beyond the occulted PSF core. The PIAA architecture reshapes and does not attenuate the incoming beam, thus all the planet light is propagated through the coronagraph and the design is highly throughput-efficient. Compared to other coronagraph architectures, this increased efficiency allows using a relatively smaller telescope diameter for the equivalent science of a telescope with a larger diameter, and is the main driver for the small-scale size of the EXCEDE mission. A set of inverse PIAA optics introduces a reverse mapping of the pupil in order to recover an undistorted field of view \cite{guyon03} that is required for imaging circumstellar material in Kuiper Belt analog regions in debris systems. 

Second, there are manufacturing limitations on the quality of all optical surfaces. Such static and quasi-static mid-spatial frequency wavefront aberration introduce starlight leakage in the form of speckles in the image plane, worsening the contrast ratio. The wavefront control (WFC) system is schematically shown in Figure \ref{fig:excedeSSS} denoted with orange and green. The WFC hardware consists of the science camera and a Deformable Mirror (DM). Using focal-plane sensing at the science camera, wavefront errors can be estimated and fed to the Deformable Mirror (DM) forming a WFC loop to correct wavefront errors and create a region of high-contrast in the image plane called the \emph{dark hole}. Finally, environmental disturbances and instabilities on the spacecraft for example due to internal vibrations and temperature gradients are measured and controlled by the Low Order Wavefront Sensor (LOWFS) system shown in purple, orange, and yellow. A high-frequency camera uses the PSF core reflected from the focal plane mask to estimate the low-order modes with tip/tilt and corrected by a fast-steering mirror (FSM) and defocus corrected by the DM. This represents a second control loop that stabilizes the wavefront operating at a higher frequency than the speckle removal loop. Finally, the resulting dark hole in the image plane is formed on the science camera's CCD detector.

In this paper, we report on the laboratory investigation that was carried out in a vacuum chamber facility at the Lockheed Martin Advanced Technology Center (described in detail in Section \ref{sect:labConf}) as part of the technology maturation program for the EXCEDE mission concept.  The main contributions presented in this paper are:
\begin{itemize}
\item experimental demonstration of high-contrast imaging using the PIAA coronagraph at a very aggressive inner working angle of 1.2 $\lambda$/D in 10\% broadband light for a test bench operating in a flight-like configuration.  More specifically, this laboratory demonstration reports $1.0 \times 10^{-5}$ raw median contrast per pixel between 1.2 to 2.0 $\lambda$/D simultaneously with $3.0 \times 10^{-7}$ median contrast between 2.0 to 11.0 $\lambda$/D. Section \ref{sect:expResults} of this paper describes the experimental data collection over three independent runs and summarizes the results that constitute a stable and repeatable contrast performance.
\item a comparison of the experimental results to the testbench optical model through a sensitivity analysis in Section \ref{sect:sensAnalysis}. We use a Gerchberg-Saxton algorithm to obtain representative aberrations from an unocculted PSF. We then demonstrate via simulation that compared to an ideal system with only $\lambda/20$ RMS phase errors, contrast in the at small angular separations is limited by low-order aberrations (first 30 Zernike terms), while contrast in the outer working zone is limited by the remaining mid-spatial frequencies (Zernike terms greater than order 30).
\end{itemize}
We first review relevant previous and existing high-contrast experiments using PIAA or other related coronagraph architectures (Section \ref{sect:background}) and compare with our experimental results -- the 1.2 $\lambda$/D inner working angle is the smallest experimentally demonstrated to date to our knowledge. Finally, we provide concluding remarks summarizing our understanding of the limiting factors in the experiment and steps suggested to further improve contrast performance (Section  \ref{sect:conc}).

\section{Background on Space Telescope Demonstrations} \label{sect:background}

To place the current experimental investigation in context, we provide a comparison with experiments on PIAA leading to the EXCEDE demosntration. We also compare results on the WFIRST-AFTA space telescope demonstrator testbeds that have been completed or are on-going (and driven by different science goals for other telescope platforms). 

A coronagraph instrument on the WFIRST-AFTA mission is currently under study and accelerated technological development. A Phase-Induced Amplitude Apodization Complex-Mask Coronagraph (PIAACMC) \cite{Guyon13} has been selected as a backup option to the primary Occulting Mask Coronagraph (a dual-mode Shaped Pupil and Hybrid Lyot that share the same architecture). The EXCEDE PIAA coronagraph and SSS technology maturation program for the EXCEDE mission concept has direct relevance for the WFIRST-AFTA backup PIAACMC, and is complementary with the on-going experimental demonstrations of the Occulting Mask Coronagraph. Whereas the WFIRST-AFTA coronagraph is expected to achieve a deeper contrast, the EXCEDE coronagraph would operate simultaneously at both a smaller inner working angle and wider outer working angle. 

The first PIAA coronagraph system laboratory validation took place at the Subaru Telescope \cite{guyon10} -- this investigation reported a $2.3 \times 10^{-7}$ raw mean contrast for a single-sided dark hole between 1.65 to 4.4 $\lambda$/D operating in air with monochromatic light and without inverse PIAA mirrors. Subsequently, a PIAA lens system was used in a temperature-stabilized testbed in air at the NASA Ames Coronagraph Experiment \cite{Belikov10} resulting in $5.4 \times 10^{-8}$ monochromatic raw mean contrast from 2.0 to 5.2 $\lambda$/D. The best raw contrast achieved using a PIAA system were obtained on the same testbed with reflective PIAA mirrors (also without an inverse PIAA system) with $1.9 \times 10^{-8}$ raw mean contrast from 2.0 to 3.4 $\lambda$/D and $1.2 \times 10^{-6}$ raw mean contrast from 1.5 to 2.0 $\lambda$/D in monochromatic light \cite{Belikov11}. Most recently, a PIAA coronagraph was used at the NASA JPL High-Contrast Imaging Testbed in 10\% broadband light from 2.0 to 4.0 $\lambda$/D recording $10^{-8}$ raw mean contrast \cite{Guyon14}.

Other coronagraph systems have achieved deeper contrasts but at larger inner working angles. As part of the WFIRST-AFTA technology demonstration, the Shaped Pupil Coronagraph testbed has recently reached its monochromatic milestone reporting $6.0 \times 10^{-9}$ raw mean contrast from 4.4 to 11.2 $\lambda$/D including the obscured AFTA pupil \cite{Cady15}.  Similarly, the Hybrid Lyot Coronagraph narrowband experimental demonstration obtained $8 \times 10^{-9}$ raw mean contrast from 3.0 to 9.0 $\lambda$/D in a 360 degree dark hole for the AFTA pupil \cite{Seo15}. Numerical studies have shown that the PIAACMC coronagraph mode with the AFTA pupil has the potential for an increase in science yield due to its smaller inner working angle and increased throughput \cite{Kern15}, but it can be sensitive to low-order aberrations (induced by line-of-sight jitter) \cite{Krist15}. Complex phase masks have only been recently manufactured \cite{Bala15} and off-axis PIAACMC mirrors have been designed \cite{Pluzhnik15}, with manufacture and laboratory investigations planned in the future. 

We emphasize that the results presented in this paper represent the smallest inner working angle demonstration to date (operating down to 1.2 $\lambda$/D), are conducted up to 10\% broadband light, with inverse PIAA optics creating a field of view out to 11 $\lambda$/D, and represent an architecture very similar to the proposed EXCEDE flight system.

\section{Laboratory Configuration}
\label{sect:labConf}

The laboratory investigation we present in this paper is the final in a series of demonstrations that were all part of a NASA-sponsored EXCEDE technology maturation process for the PIAA-based starlight suppression system. The earlier technology demonstrations \cite{excedeWhite1, excedeWhite2} were first carried out in air then the temperature-stabilized testbed at the NASA Ames Coronagraph Experiment using only forward PIAA mirrors \cite{Belikov12, Belikov13}, and subsequently incorporated the first operation of the fast Low Order Wavefront Sensor \cite{Lozi13, Lozi14}. In this paper, we will focus exclusively on high-contrast results obtained in a vacuum environment with an optical bench that represents the final EXCEDE-like flight configuration \cite{Belikov14, Sirbu15}.

\begin{figure}[t!]
\centering
\subfloat[]{
\centering
\label{fig:testbed-entrance}
\includegraphics[height = 2.35 in]{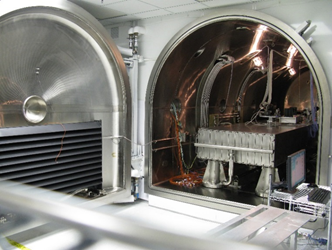}
}
\subfloat[]{
\centering
\label{fig:testbed-camera}
\includegraphics[height = 2.35 in]{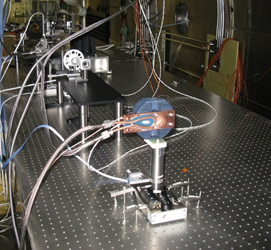}
}\\
\subfloat[]{
\centering
\label{fig:testbed-forwardPIAA}
\includegraphics[height = 2.125 in]{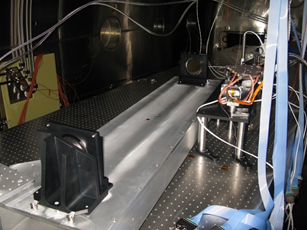}
}
\subfloat[]{
\centering
\label{fig:testbed-inversePIAA}
\includegraphics[height = 2.125 in]{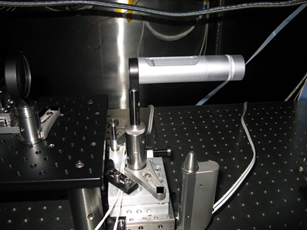}
}\\
\subfloat[]{
\centering
\label{fig:testbed-frontEnd}
\includegraphics[height = 2.125 in]{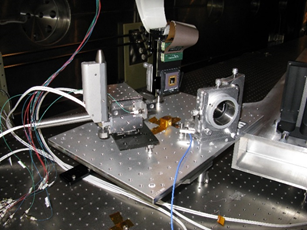}
}
\subfloat[]{
\centering
\label{fig:testbed-focalPlaneMask}
\includegraphics[height = 2.125 in]{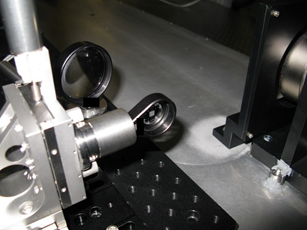}
}
\caption[Vacuum chamber and optical testbench.]{\label{fig:testbed} Experimental demonstration was performed in the vacuum MET chamber facility at the Lockheed Martin Advanced Technology Center. \subref{fig:testbed-entrance} Class 10,000 clean room at the entrance of the vacuum chamber. \subref{fig:testbed-camera} View of the EXCEDE optical testbed from the science camera end. \subref{fig:testbed-forwardPIAA} View of forward PIAA mirrors.  \subref{fig:testbed-inversePIAA} View of inverse PIAA lenses. \subref{fig:testbed-frontEnd} View of front-end optics on platform with two OAPs and the DM. \subref{fig:testbed-focalPlaneMask} View of the focal plane mask}
\end{figure}

\subsection{Experiment Facility}

The EXCEDE laboratory demonstrations were carried out in a vacuum chamber facility at the Lockheed Martin Advanced Technology Center. This vacuum ‘Metrology’ (MET) chamber is embedded in a 20 ft wide by 40 ft long (6 m by 12 m) class 10,000 cleanroom. The chamber itself is 8 ft (2.4 m) wide and sufficiently tall to allow comfortable access in air to the entire length of the 4 x 20 x 2 ft (1.2 x 6.1 x 0.6 m) optical table that is integrated to the chamber through vibration isolators located external to the chamber. The entrance to the chamber is shown in Figure \ref{fig:testbed}\subref{fig:testbed-entrance} and the final configured testbed are shown in Figures \ref{fig:testbed}\subref{fig:testbed-camera}-\subref{fig:testbed-focalPlaneMask}. 

The testbed features no enclosure other than the chamber itself, as the intent is to operate in vacuum. This implies, however, that high-contrast cannot be realistically achieved in air (while the chamber is open) due to an uncontrolled environment. The vacuum environment within the chamber is produced by an Edwards dry roughing pump, a cryogenic pump, a turbo molecular pump, and a liquid nitrogen getter plate. In vacuum, thermal gradients are monitored and mapped (but not controlled) with temperature probes located on important components throughout the testbed, and the cameras cooled and outfitted with heat-sinks. Environmental data consisting of pressure, temperature, vibration, and acoustic measurements are logged, and these data were analyzed in more detail as part of the presentation of the initial vacuum results \cite{Belikov14}.

The vacuum testbed was incrementally configured toward our final milestone demonstration over the course of five separate Vacuum Chamber Tests (VCTs) from December 2013 to March 2015 to as much as possible closely resemble the proposed EXCEDE configuration. The primary considerations for this progressive configuration were the following:
\begin{enumerate}
\item enabling an exploration/demonstration of inverse PIAA-corrected contrast performance with broadband light with up to a 10\% bandwidth.
\item verification of a wide outer working zone, with the final verification of the outer working angle set to 11 $\lambda$/D. 
\item demonstration of system operability in a vacuum environment.
\end{enumerate}
In practical terms, the main laboratory features that reflect these requirements are: (i) the usage of a supercontinuum white laser source with a selectable bandwidth filter for polychromatic contrast, with reflective optics used to mitigate chromatic effects and refractive optics used for less chromatically-sensitive functions, (ii) introduction of inverse PIAA optics to enable a wider field of view with an undistorted, off-axis PSF in the final focal plane, (iii) a DM position upstream of the forward PIAA optics, to work in conjunction with the inverse PIAA system and to enable removal of tip/tilt modes with sensing provided by the LOWFS prior to propagation through the forward PIAA system, and (iv) preparation of electronic components and optical mounts for the testbed for vacuum compatibility. 

\subsection{Optical Testbed} \label{sec:testbedDescription}

The final optical configuration of the EXCEDE testbed for polychromatic demonstration in vacuum is shown schematically in Figure \ref{fig:excedeConfiguration}. The propagation of the beam from the external source input to the science camera imaging detector at the final focal plane is described below:

\begin{figure}[t!]
\centering
\includegraphics[height = 2.0 in]{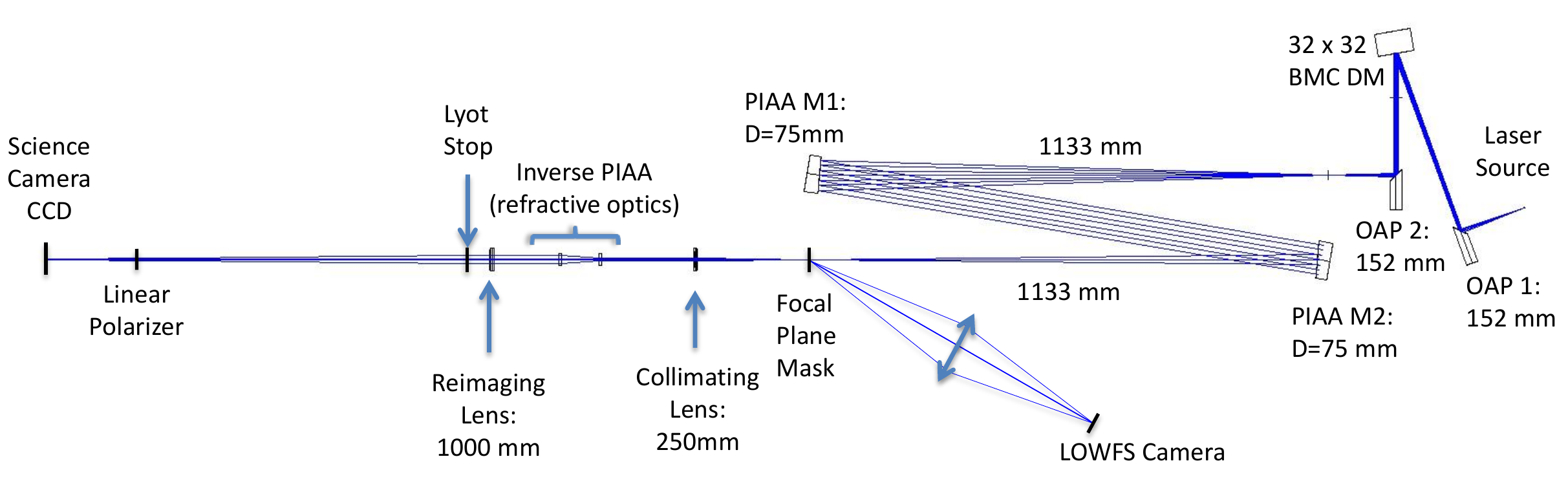}
\caption[Optical configuration of the EXCEDE test bench.]{\label{fig:excedeConfiguration} Optical configuration of the EXCEDE test bench.}
\end{figure}

\textbf{LASER SOURCE}. The laser source used is a supercontinuum light source (an NKT photonics SuperK laser) with a variable bandwidth filter (the SuperK VARIA tunable single line filter) coupled into a single mode fiber. The single-mode fiber output is an approximation to a point-source (stellar) image delivered by a telescope, and is fed into the vacuum chamber via a port hole. The central wavelength was selected to be 650 nm due to being in the optical region and as approximately mid-way between the two proposed EXCEDE flight-configuration spectral passbands. All the optics upstream from the focal plane mask are reflective to mitigate chromatic effects. Refractive optics are used after the focal plane mask to ensure their chromatic aberrations do not impact the ability of the focal plane mask to suppress the on-axis point-source PSF and, additionally, are all placed in a slow beam (f/67 or slower). 

\textbf{OAPs}. The laser source feeds front-end optics containing the DM that was positioned upstream of the forward PIAA system conforming to the EXCEDE flight architecture. The front-end optics are shown in Figure \ref{fig:testbed}\subref{fig:testbed-frontEnd}. Two off-axis parabolic mirrors (OAP)s, with $\lambda$/4 surface quality, are used to: (a) create a point focus input to the forward PIAA with an f/15 beam, and (b) conjugate the DM with respect to the first forward PIAA mirror (M1).

\textbf{DM}. The DM used is a 32x32 actuators Boston Micromachines (BMC) MEMS. The DM is conjugated to PIAA M1 to avoid geometrical distortion effects due to the forward PIAA system and allow for wide field correction. The DM converts commanded voltages into displacement of individual actuators. An iterative speckle-nulling (SN) wavefront correction algorithm was used to compute the DM settings creating the high-contrast dark hole in the image plane.

\textbf{FORWARD PIAA MIRRORS}. The beam next propagates to the forward PIAA mirrors shown in Figure \ref{fig:testbed}\subref{fig:testbed-forwardPIAA}. These are the same first-generation Axsys PIAA mirrors used in the Subaru \cite{guyon10}, JPL \cite{Guyon14}, and earlier EXCEDE results \cite{Belikov13, Belikov14}. They were not specifically optimized for broadband performance nor a 1.2 $\lambda$/D focal plane mask. 

\textbf{FOCAL PLANE MASK}. The  focal plane mask (FPM), with a photograph shown in Figure \ref{fig:testbed}\subref{fig:testbed-focalPlaneMask} and a schematic shown in Figure \ref{fig:piaa}\subref{fig:piaa-fpm}, is at the prime focus of the forward PIAA system. The focal plane mask has a C-shape deposit on glass with a reflective circular inner occulter blocking the core of the PSF. The light reflected from the inner circular occulter feeds to the LOWFS. A straight edge blocks one entire side of the focal-plane, and an outer circle blocks light at off-axis angles greater than 16 $\lambda$/D.

\textbf{LOWFS}. The central core of the PSF is reflected by the focal plane mask to the LOWFS system. A re-imaging lens is used to form a slightly defocused image on the LOWFS camera. A 14-bit ImperX Bobcat ICL-B0620 CCD camera provides a high frame rate and a National Instruments PXI controller establishes a real-time control loop with the DM. The design, performance, and operation of the EXCEDE testbed LOWFS are detailed in earlier reports \cite{Lozi13, Lozi14}. We emphasize that only the tip/tilt modes are estimated and corrected in this experiment. Additionally, the correction loop operates at 1 kHz and is closed with appropriate settings set on the DM (rather than a fast steering-mirror as proposed in the EXCEDE flight-architecture).

\textbf{INVERSE PIAA LENSES}. The inverse PIAA lenses are shown in Figure \ref{fig:testbed}\subref{fig:testbed-inversePIAA} and are placed in a collimated space in order to reduce refractive effects in broadband light. As these lenses are located after the FPM, their aberrations do not impact the ability of the FPM to suppress the on-axis light. 

\textbf{LYOT STOP}. An undersized Lyot stop (with a 0.85 fraction of the full pupil diameter) is placed shortly after a final re-imaging lens, and in nearly collimated space as the f/67 of the final lens is slow. This defines the exit pupil and blocks out the edges of the second pupil plane diffracted light. The light blocked by the Lyot stop is high-spatial frequency and is due to to the focal plane occulter's edge diffraction effects \cite{Belikov11}.

\textbf{LINEAR POLARIZER}. Earlier experiments conducted at Ames have shown that in the contrast regime that the EXCEDE experiment operates polarization effects can become a limiting factor particularly in the outer working zone \cite{Belikov11}. A linear polarizer is therefore located immediately prior to the science camera and mitigates any instrumental polarization effects due to reflective optics in the system.

\textbf{SCIENCE CAMERA}. The science camera, located in a re-imaged focal plane after the linear polarizer, is used both for target imaging, and for closed loop mid-spatial frequency wavefront sensing and control with the DM. It has an as-calibrated linear image scale of 5.5 pixels per 1 $\lambda$/D. This 16-bit QSI 520i series CCD camera provides a low read-noise output and is shown in Figure \ref{fig:testbed}\subref{fig:testbed-camera}. It has a regulated thermoelectric cooler which is maintained at 1° C to avoid ice crystal formation during operation. We discovered contaminant deposition on both the exterior vacuum window and the surfaces of the CCD window (a few mm ahead of the final focal plane) during in-air preparatory activities before the final VCT. These manifested as slightly afocal (diffractive) spots on the science camera image. Removal of the vacuum window eliminated the largest-sized diffraction spots. The contaminant on the internal surface of the CCD window resulted in remaining small nucleation spots and can be seen in the final reduced experimental figures (see the reduced images in Figure \ref{table:experimentalResults}). The presence of these small spots had no adverse effects of significance on either the performance of the SN WFC algorithm, calibration procedures, or median image contrasts (though it did affect mean contrast at the smallest inner working angles -- hence we report both mean and median contrasts).

Compared to the flight-concept EXCEDE SSS discussed in \S \ref{sect:intro}, there are some differences in the laboratory implementation. The tip/tilt removal as part of the LOWFS was implemented on the upstream DM as there was no fast-steering mirror in the testbed. Additionally, the LOWFS only corrected tip/tilt modes with no active focus control. The testbed is operating in a single visible band for this experimental demonstration. 

\subsection{Wavefront Control}

The speckle nulling (SN) wavefront control algorithm was used in this experiment due to its maturity in the Ames Coronagraph Experiment testbed \cite{Belikov10, Belikov11, Belikov12}.  Modern SN implementations follow from the original iterative energy minimization formulation in Malbet and Shao (1995). SN improves robustness of the iterative solution by reducing the area of the dark hole for which total energy is iteratively minimized to individual speckles -- this comes at the cost of increasing the total number of iterations. 

For the SN algorithm, a set of $n$ brightest speckles are identified in the science camera's focal plane subject to some constraints such as minimum separation between speckles. SN constraints determine the aggressiveness of the SN algorithm -- for example smaller applied DM strokes and  larger separation between corrected speckles reduces correlation between speckles being simultaneously corrected. The speckle positions in the focal plane can be mapped to the necessary spatial frequency on the DM. Each speckle is then interfered with a DM probe which scans in both phase and amplitude. The phase and amplitude of the DM spatial frequency that minimizes the energy contained within each speckle are chosen for that iteration's DM settings.

The closest implementation of SN in literature to the one in use on the EXCEDE bench is from the Subaru's SCEAxO \cite{Martinache12}. The main difference with the implementation of speckle nulling utilized in the EXCEDE testbed is that we use both amplitude and phase. Additionally, the EXCEDE bench is operating at a deeper contrast level than the on-sky SCEAxO, which achieves an average contrast level of $3 \times 10^{-4}$ in the control region, and requiring more iterations (on the order of a thousand iterations for the EXCEDE bench). To reach these deeper contrast levels, our SN strategy is to set aggressive settings for early iterations. These include a larger number of speckles (10-20 speckles at a time) with a larger radius in which energy is minimized and a more sparsely sampled parameter space for the phase and amplitude probes. As deeper contrast levels are attained, the SN setting aggressiveness is reduced in order to minimize speckle interactions during probing. When a contrast plateau is reached, the SN aggressiveness may be increased for a number of iterations before again reducing the settings aggressiveness. Typically, by the end of the SN loop only a single speckle is corrected per iteration at small angular working separations.

Model-based iterative techniques such as Electric Field Conjugation (EFC) \cite{Giveon07} can greatly reduce the total number of iterations; however, performance does depend significantly on the accuracy of the model used and therefore can be limited by these model errors. Given a testbed configuration that provides sufficient stability for the speckle nulling implementation to converge, the algorithm is not expected to be a limiting factor at the EXCEDE contrast levels for wavefront probing in monochromatic light. One limitation of SN is that speckle probing works best for monochromatic light as the phase and amplitude estimate are degraded by speckle smearing; on the EXCEDE testbed we found that applying speckle probing directly in broadband light results in contrast levels one order of magnitude worse than reported in this study. Estimation in direct broadband light requires more advanced model-based techniques \cite{Sirbu15b}.

\subsection{System Calibration}

In order to ensure repeatability and accuracy of the system, the EXCEDE bench uses automated calibration procedures to align the various components of the system \cite{excedeWhite3}. This is an important part of the experimental procedure and useful for vacuum operation when the testbench hardware cannot be accessed. Operation of the bench is run via a LabVIEW GUI, which is able to launch different procedures which align the different parts of the system and calibrate the system model; when all the alignment procedures have been completed it is possible to start the iterative WFC algorithm. Here we summarize the main calibration procedures used on the bench during each demonstration run:


\textbf{INPUT FIBER ALIGNMENT}: The input fiber is aligned in both orthogonal directions (X, Y) in the transverse plane with respect to the optical axis, to assure on-axis concentricity with respect to both the front-end optics, and to the PIAA system. Due to the inverse PIAA optics, the science camera cannot be used for this alignment. Instead, the PSF core reflected light from the FPM is imaged on the LOWFS camera. A sharpness metric defined as $\Sigma =\textrm{image}^2 / (\Sigma \textrm{image})^2$ over all PSF pixels on the LOWFS camera is maximized while the fiber is moved both along the X- and Y-axes. The LOWFS camera is set at a slightly defocused location to obviate a trans-focal parity degeneracy \cite{Guyon09}. The procedure is iterative and run until the optimal location is stable.

\textbf{INVERSE PIAA ALIGNMENT AND IMAGE SCALE}: The inverse PIAA lenses must be separately aligned to ensure on-axis concentricity with respect to the forward PIAA mirrors and the front-end optics. Additionally, the best-focus location for the science camera is established, and the image scale in terms of pixels per $\lambda$/D at the science camera image plane is measured. The FPM is moved out of the way to ensure the unocculted PSF is visible on the science camera. The inverse PIAA lens assembly is moved along both transverse axes (X, Y), and the maximal sharpness locations are maintained. The science camera is also moved along the optical axes (Z) to ensure the best focus. The fiber is displaced a known amount along both X and Y to determine the final image plane scale. 

\textbf{CONTRAST CALIBRATION}: An unocculted reference image of the PSF of the aligned system is obtained at the same central wavelength used for the Wavefront Control iteration algorithm. The FPM is moved to allow the PSF to pass through a transparent part of the mask. To obtain contrast-calibrated images, every pixel in the final focal plane is divided by a calibration factor which is peak flux density obtained from the unocculted, unsaturated PSF. For broadband contrast images, the science camera saturates at the shortest possible available exposure times (and reducing the overall intensity would significantly increase the duration of each wavefront control iteration). We therefore used integrated measurements of the flux from the reflected core light on the LOWFS camera for both the monochromatic and broadband PSFs to obtain the necessary calibration ratio. 

\textbf{FPM ALIGNMENT}: To ensure that the C-shaped focal plane occulter mask was correctly aligned and positioned the FPM was moved along both transverse axes (X, Y). At each location the total energy in the image obtained was summed, and divided with respect to the energy in the unocculted PSF. Along the Y axis (coinciding with the mask’s straight edge), the relative energy was minimized; along the X axis the FPM location was chosen such that the relative energy was 50\%.

\textbf{IWA VERIFICATION}: A separate procedure was used to verify the location of the C-shaped occulter without the necessity of re-alignment of the occulter (i.e., to ensure that hysteresis does not result in a different FPM location). The fiber was displaced along the X axis by known amounts and the relative energy was measured. This process assures that the PSF was not moved by tilts introduced on the DM by the wavefront control algorithm, and that no other changes in the system affected the location of the IWA. The verification procedure has $\pm 0.05 \lambda$/D uncertainty.

\textbf{LOWFS CALIBRATION}: The LOWFS uses reflected light from the focal plane mask. The core of the reflected PSF from the circular inner part of the C-shape mask is focused onto a high frame-rate camera. For calibration, a reference image of the reflected PSF is constructed by collecting an initial set of images from the LOWFS. Subsequent images are subtracted with respect to the reference image and the shifted centroid is computed. A known set of tip/tilt modes are applied to the DM to construct an influence matrix with respect to the LOWFS focal plane. 

\section{Experimental Results}
\label{sect:expResults}

\begin{figure}[t!]
\centering
\includegraphics[height = 3.65 in]{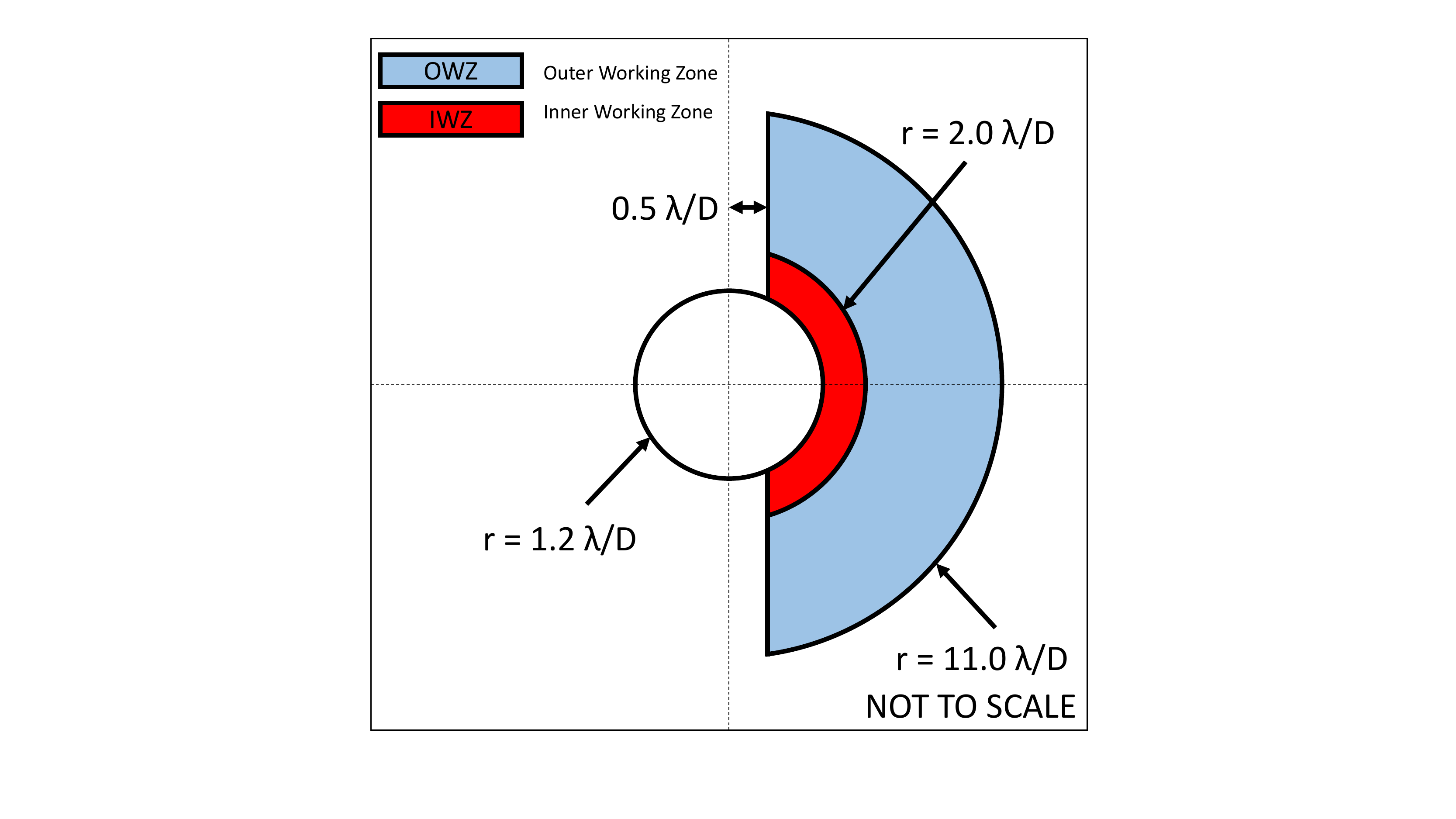}
\caption[Definition of the single-sided dark hole, its inner working zone (IWZ), and outer working zone (OWZ).]{\label{fig:darkhole} Dark hole definitions for EXCEDE measurements. Inner working zone (IWZ) is defined between 1.2 and 2.0 $\lambda$/D and Outer Working Zone (OWZ) is between 2.0 and 11.0 $\lambda$/D.}
\end{figure}

In this section, we report on the demonstration of broadband contrast in the vacuum chamber carried out as part of the final VCT. The geometry of the dark hole wherein contrast measures are obtained is illustrated in Figure \ref{fig:darkhole}, with an Inner Working Zone (IWZ) defined between 1.2 and 2.0 $\lambda$/D and Outer Working Zone (OWZ) between 2.0 and 11.0 $\lambda$/D. The experimental measurements were demonstrated in a precisely defined and repeatable fashion. To demonstrate repeatability the full set of experimental measurements is taken three times. To ensure independence of each trial, the bench is recalibrated via the automated LabVIEW procedures and DM settings are set to the initial flat prior to starting each demonstration run. To demonstrate stability (within each run) and repeatability (from run to run), at least 1,000 individual frames were collected over an interval of at least approximately 1 hour in each independent run. The worst 10\% frames were rejected (as per on-orbit processing design) and the median per pixel contrast was computed from the remaining 90\% frames.

\begin{figure}[t!]
\centering
\includegraphics[height = 3.65 in]{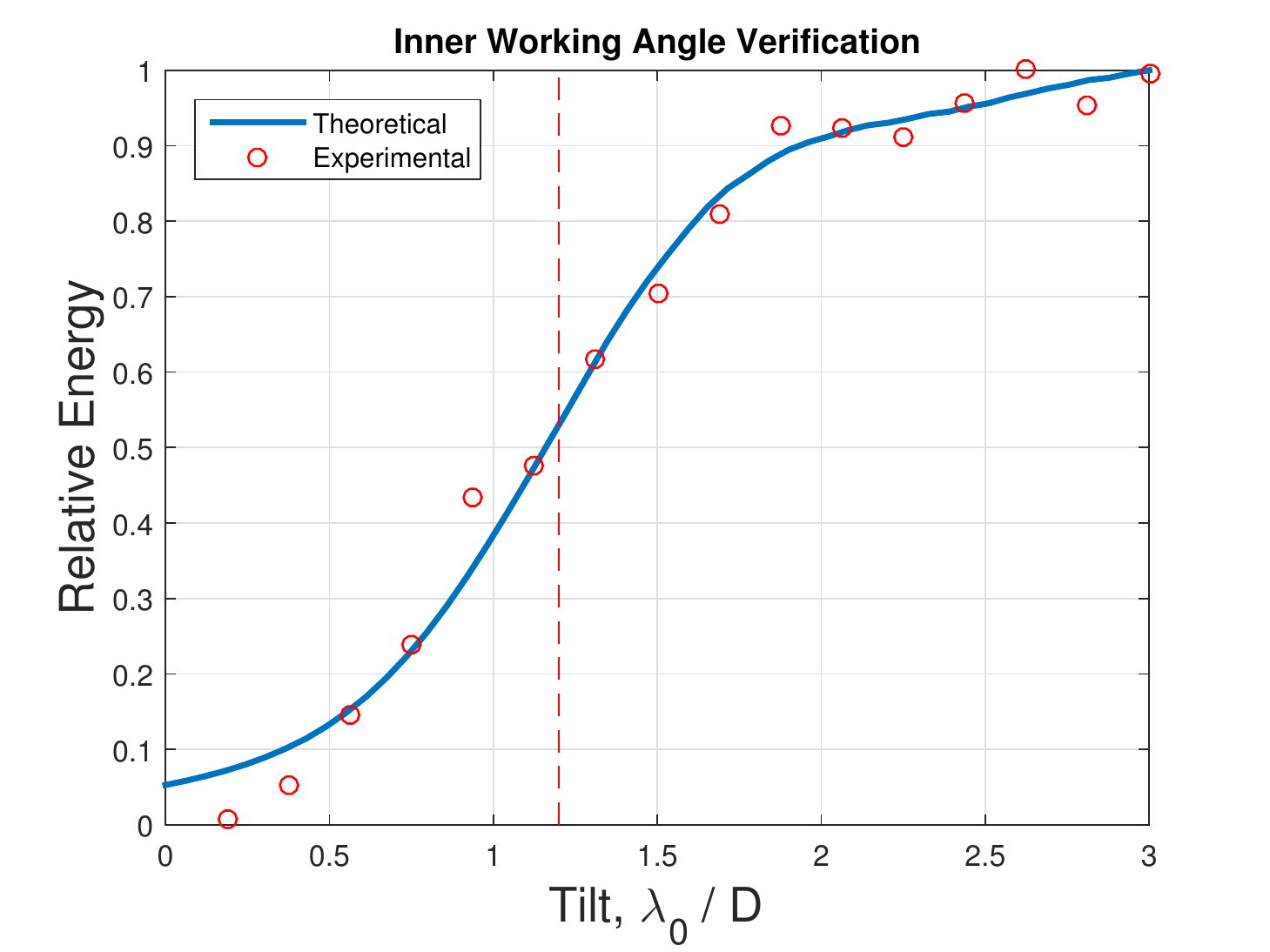}
\caption[Inner working angle verification routine]{\label{fig:iwa_verification} Comparison of simulation and experimental verification of the inner working angle occurrence (defined as 50 \% throughput) at 1.2 $\lambda/D$.}
\end{figure}

The demonstration is performed in 10\% bandwidth (65 nm) centered about 650 nm. Initially, the system calibration and Speckle Nulling (SN) WFC iterations are performed at the smallest selectable filter bandwidth available on the VARIA (10 nm). The LOWFS is engaged once the IWZ contrast limit is reached -- this particular limit is around $1 \times 10^{-5}$ in the 1.5\% (10 nm) narrowband wavelength, and without the LOWFS a IWZ contrast better than $1 \times 10^{-5}$ cannot be reached. Note that before launching the LOWFS, we recalibrate the position of the mask to ensure that the 50\% throughput is located at 1.2 $\lambda$/D after the shape of the PSF has stabilized. The final IWZ contrast achieved in 10 \si{\nano \meter} is ~$3 \times 10^{-6}$. Once contrast has stabilized, we increase the bandwidth to 65 \si{\nano \meter} (i.e., 10\% bandwidth) and measure the resulting broadband contrast. At the end of the wavefront control run, we verify the location of the inner working angle (defined to be the 50 \% throughput line) is maintained at 1.2 $\lambda$/D -- the experimental measured normalized energy is shown in Figure \ref{fig:iwa_verification} against the theoretical demonstrating that the WFC iterations did not introduce a PSF tilt. The results from each of the three separate runs are summarized below in Figure \ref{table:experimentalResults}.
\begin{figure}[h!]
\begin{center}
\begin{tabular}{| c | c | c |}
\hline
\multirow{2}{*}{Summary} & Image & Stability over  \\
 & (Raw Contrast) & 1000 $\geq$ iterations \\
\hline 
& & \\
\parbox[t]{5cm}{\emph{Test A} \\ Time interval: 67 mins \\ Mask position: 1.2 $\lambda$/D \\ $\lambda_\textrm{central}$ = 650 \si{\nano \meter} \\ Bandwidth = 10\% \\ \vskip1pt Median raw contrast:\\ 1.2-2.0 $\lambda$/D: $1.35 \times 10^{-5}$ \\ 2.0-11~ $\lambda$/D: $2.82 \times 10^{-7}$} & {\raisebox{-\totalheight}{\includegraphics[height = 1.8 in]{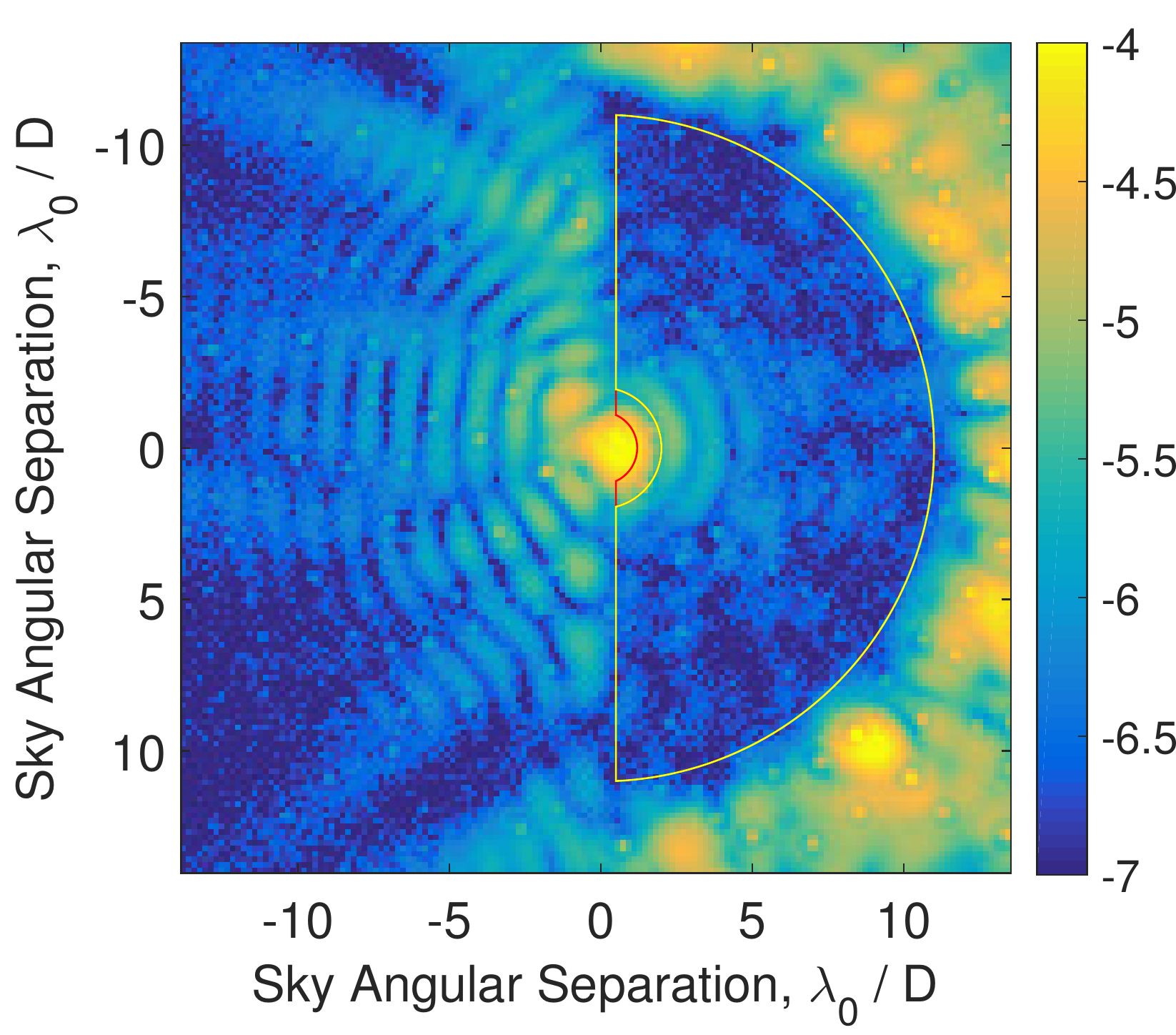}}} & \multirow{2}{*}{\raisebox{-\totalheight}{\includegraphics[height = 1.8 in]{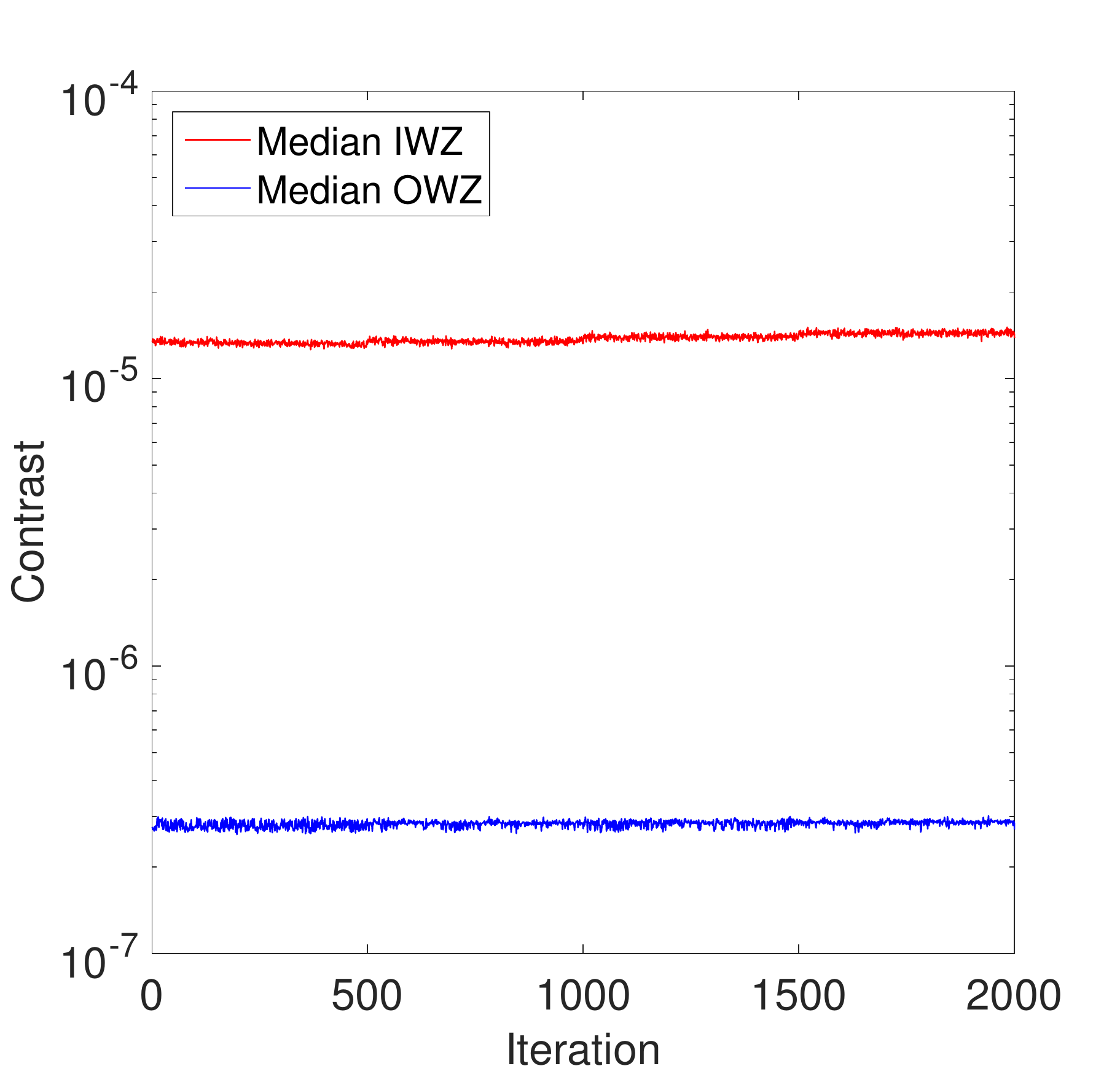}}} \\
\hline
\parbox[t]{5cm}{\emph{Test B} \\ Time interval: 816 mins \\ Mask position: 1.2 $\lambda$/D \\ $\lambda_\textrm{central}$ = 650 \si{\nano \meter} \\ Bandwidth = 10\% \\ \vskip1pt Median raw contrast:\\ 1.2-2.0 $\lambda$/D: $1.29 \times 10^{-5}$ \\ 2.0-11~ $\lambda$/D: $3.14 \times 10^{-7}$}  & \raisebox{-\totalheight}{\includegraphics[height = 1.8 in]{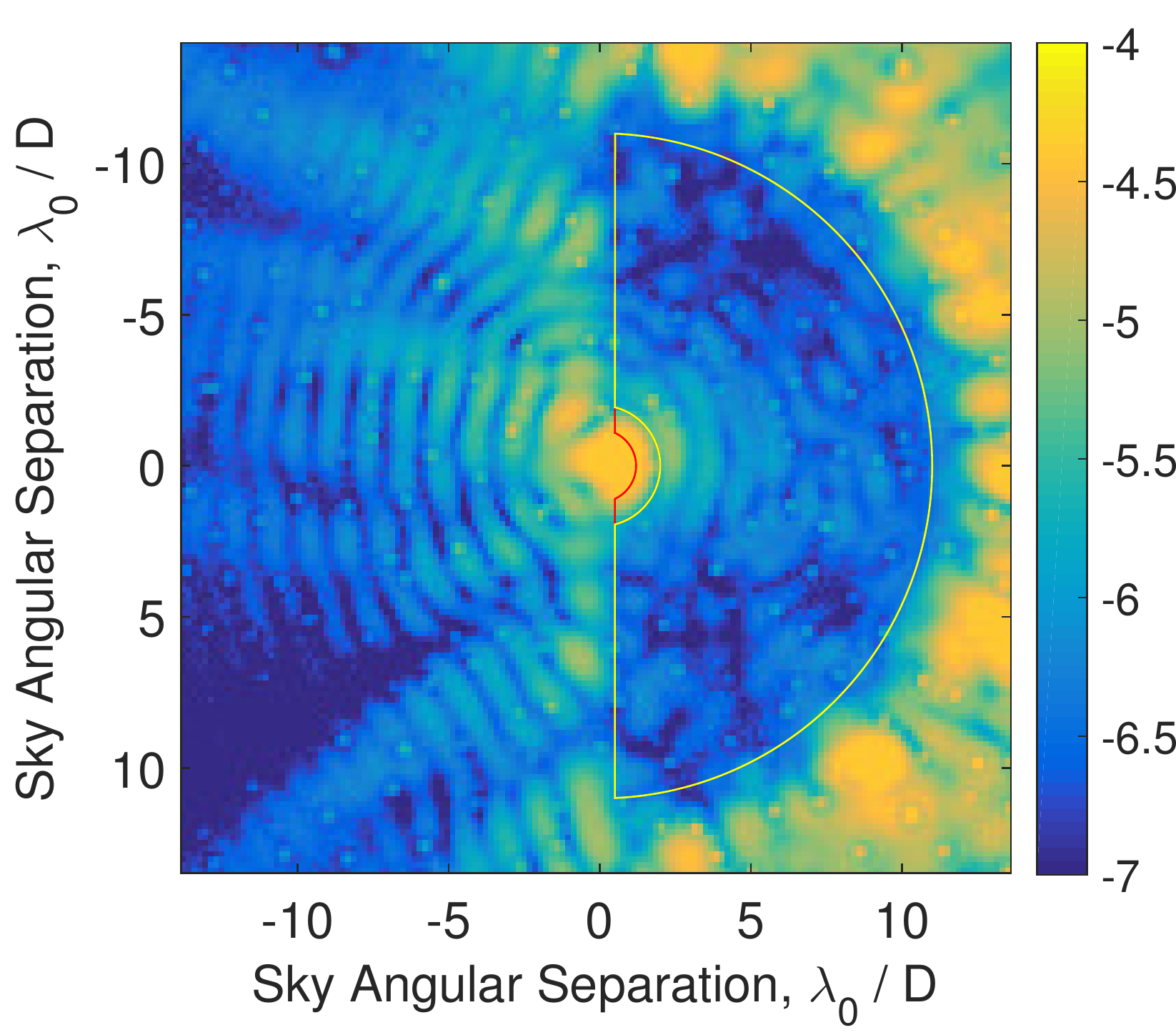}} & \raisebox{-\totalheight}{\includegraphics[height = 1.8 in]{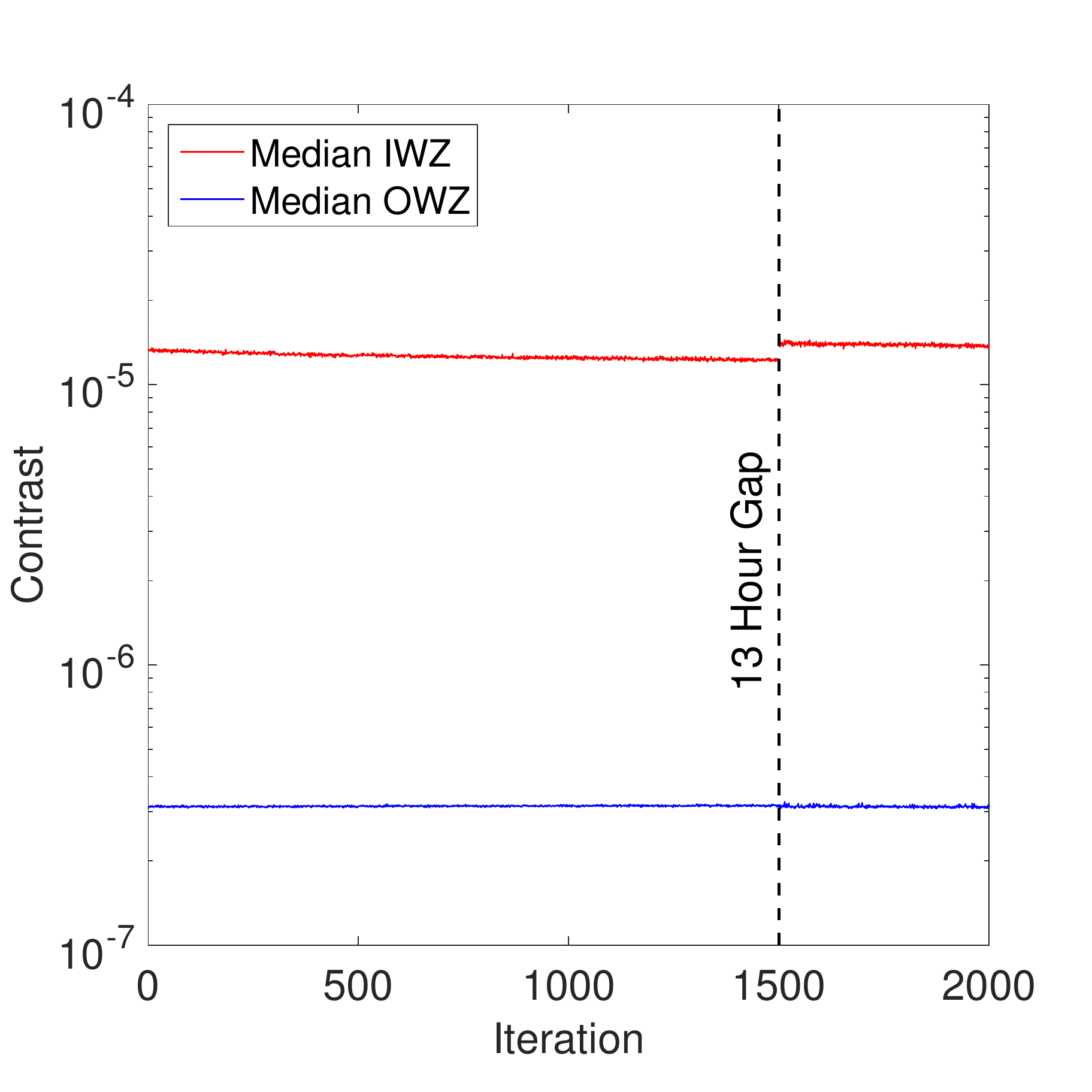}} \\
\hline
\parbox[t]{5cm}{\emph{Test C} \\ Time interval: 61 mins \\ Mask position: 1.2 $\lambda$/D \\ $\lambda_\textrm{central}$ = 650 \si{\nano \meter} \\ Bandwidth = 10\% \\ \vskip1pt Median raw contrast:\\ 1.2-2.0 $\lambda$/D: 
$1.33 \times 10^{-5}$ \\ 2.0-11~ $\lambda$/D: $2.63 \times 10^{-7}$}& \raisebox{-\totalheight}{\includegraphics[height = 1.8 in]{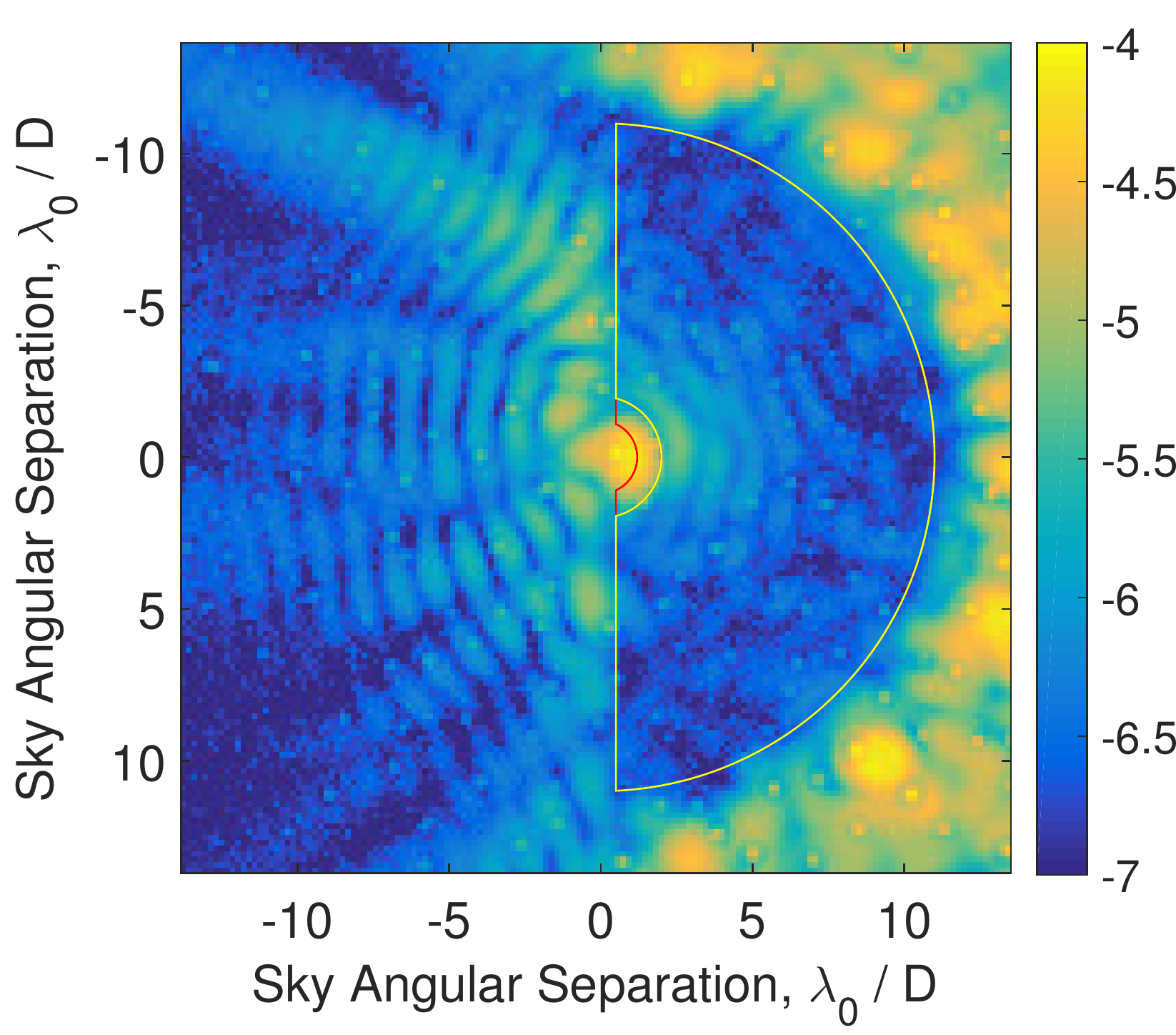}} & \raisebox{-\totalheight}{\includegraphics[height = 1.8 in]{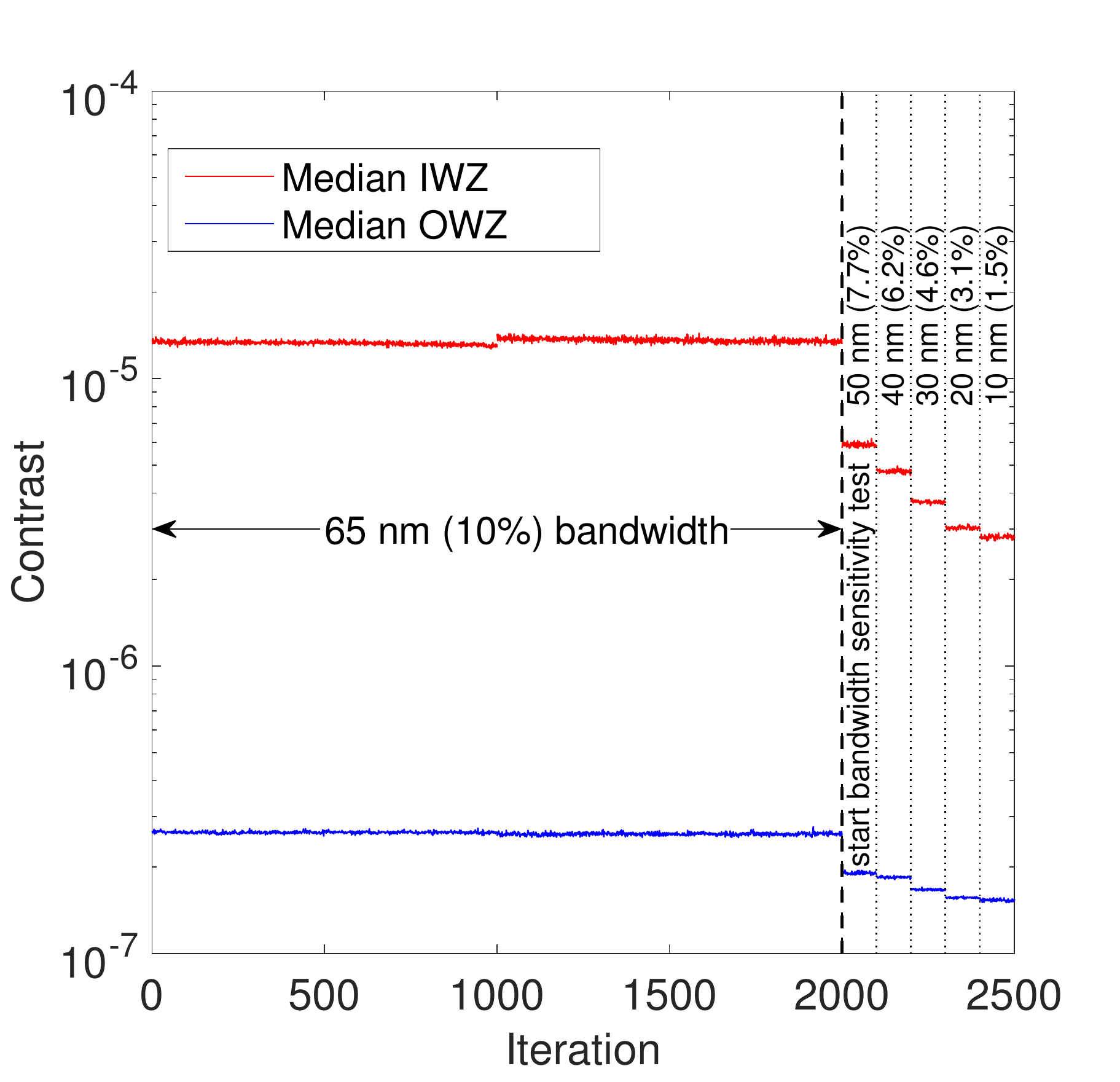}} \\
\hline
\end{tabular}
\end{center}
\caption[Summary of the three experimental runs and contrast results.]{\label{table:experimentalResults} Summary of the three experimental runs for the demonstration of 10\% bandwidth starlight-suppression: 1.2 to 11 $\lambda$/D at 650 \si{\nano \meter}. The small nucleated spots on the experimental images are contaminant deposition on the inside of the science camera CCD window from outgassing in the vacuum environment.}
\end{figure}

For each of the three test runs characterized in Figure \ref{table:experimentalResults} a single, but typically representative, contrast field image (from the $\geq$ 1000 obtained) is shown under the Image column. The individual images are highly repeatable, as evidenced in the contrast stability plots shown under the Stability column for each test presenting the raw median contrast metric as a function of image iteration. In all three cases, data were collected with the establishment of the dark zone by SN for $\geq$ 1000 iterations, as follows:

\begin{itemize}
\item In Test A (top panels), 2,000 images were collected contiguously with a time-averaged cadence of one frame every 2 seconds (over 67 minutes).
\item In Test B (middle panels), we took the opportunity to test the longer term stability of SN by continuously taking data over $\sim$ 14 hours. We began with 1,500 images collected similarly to Test A. Then, without resetting the DM, we explored broadband performance at other bandwidths (from 50 nm to 10 nm, not illustrated here, but compared with models in Section \S \ref{sect:sensAnalysis}) before contiguously collecting another 500 images at 65 nm (corresponding to 10\% bandwidth).  The stability of the SN-established WFC over that period of time is evidenced by the contrast metrics at 10\% bandwidth that are graphed (collapsed across the 13-hour gap when data were being taken at other bandwidths) in the column on the right.
\item In Test C (bottom panels) was executed similarly to Test B, but with 2,000 images at 65nm (10\%) bandwidth taken with different exposure times (1000 images with 0.1 s exposures, followed by another 1000 iterations with 0.14 s exposures after a 20 minute interval). We use those 2,000 images for our 10\% bandwidth contrast metrics. Following that, as in test B, the bandwidth was incrementally decreased in consecutive iterations, in this case, of 100 images each that are shown here for illustrative purposes only. The 65 nm (10\%) bandwidth results are virtually identical to Tests A and B in the OWZ.
\end{itemize}

\section{Sensitivity Analysis}
\label{sect:sensAnalysis}

To better understand the physical limitations of the as-implemented testbed, and in particular to identify the factor(s) limiting the experimental performance informed by experimental test results, specifically at the smallest stellocentric angle in the IWZ between 1.2 and 2.0 $\lambda$/D, we provide here in detail a description of, and the results from a performance sensitivity analysis. 

\subsection{Model Description}

\begin{figure}[t!]
\centering
\includegraphics[height = 2.5 in]{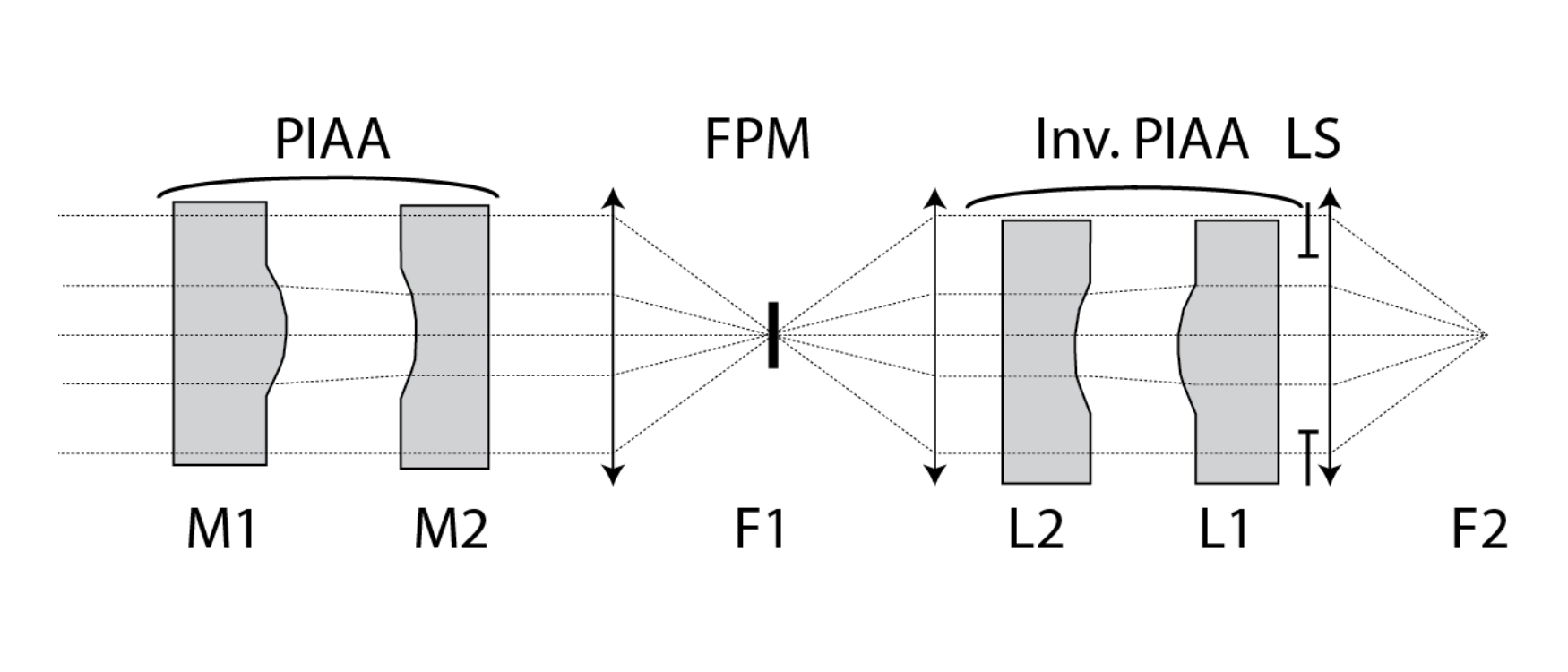}
\caption[Schematic of optical elements used.]{\label{fig:model} Optical element and plane definitions for the optical model employed. The forward PIAA consists of the two mirrors M1 and M2 (illustrated only for simplicity as transmissive optics). The Focal Plane Mask (FPM) is located at the first focal plane F1. The Inverse PIAA consists of two inverse lenses L2 and L1. A Lyot stop (LS) is located at the exit pupil, and the final science image is formed at the re-imaged focal plane F2.}
\end{figure}

We have adopted a geometrical remapping optical propagation model between the PIAA mirrors and with all the optical planes in the system modeled defined as Fourier conjugates. Despite its relative simplicity, the model replicates the observed performance (and limitations) of the experiment quite well (as we will demonstrate). This model of the EXCEDE testbed coronagraphic optics is schematically illustrated at a high level in Figure \ref{fig:model}.

\begin{figure}[t!]
\centering
\subfloat[]{
\centering
\label{fig:piaa-m2}
\includegraphics[height = 2.5 in]{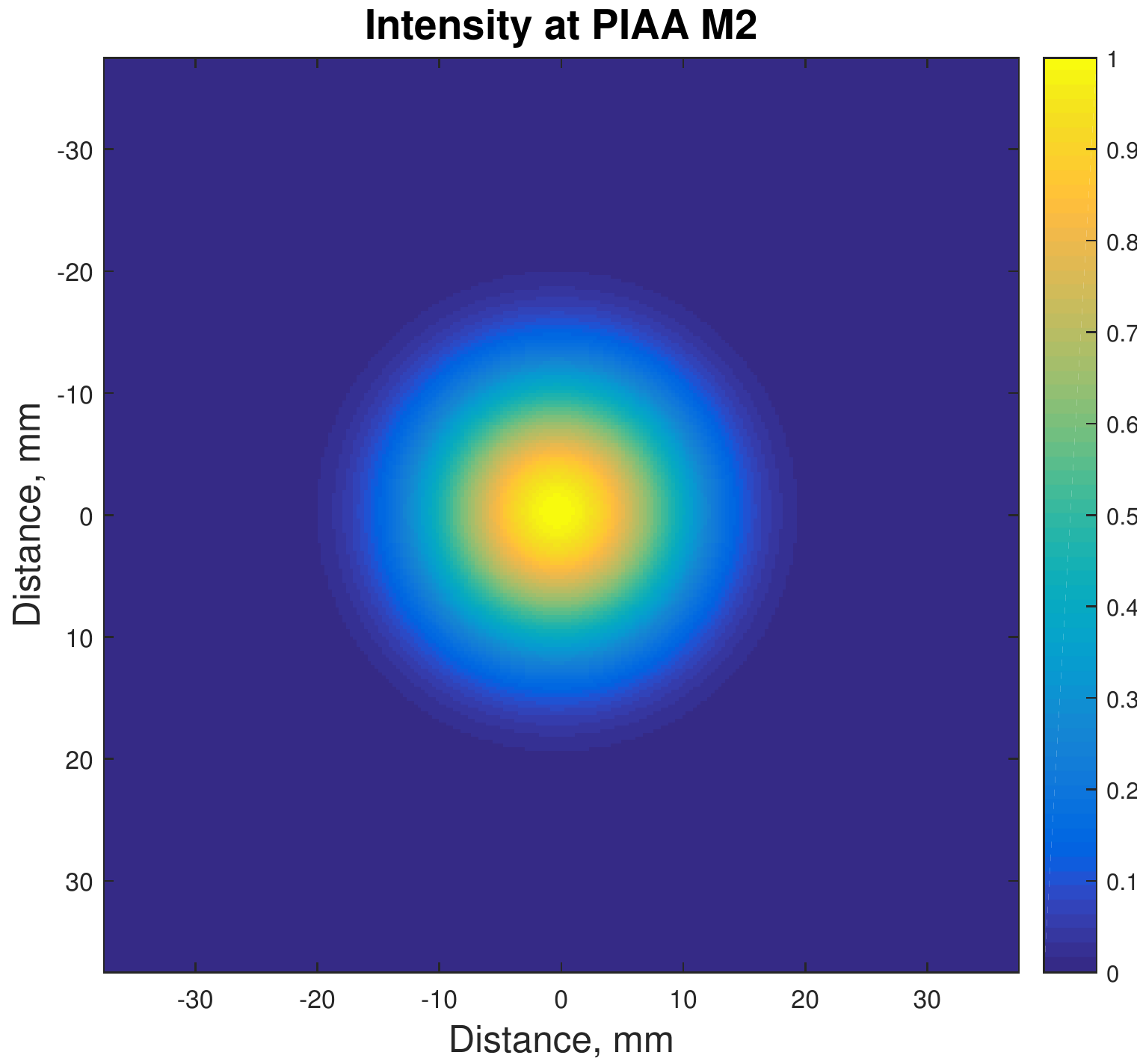}
}
\subfloat[]{
\centering
\label{fig:piaa-fpm}
\includegraphics[height = 2.5 in]{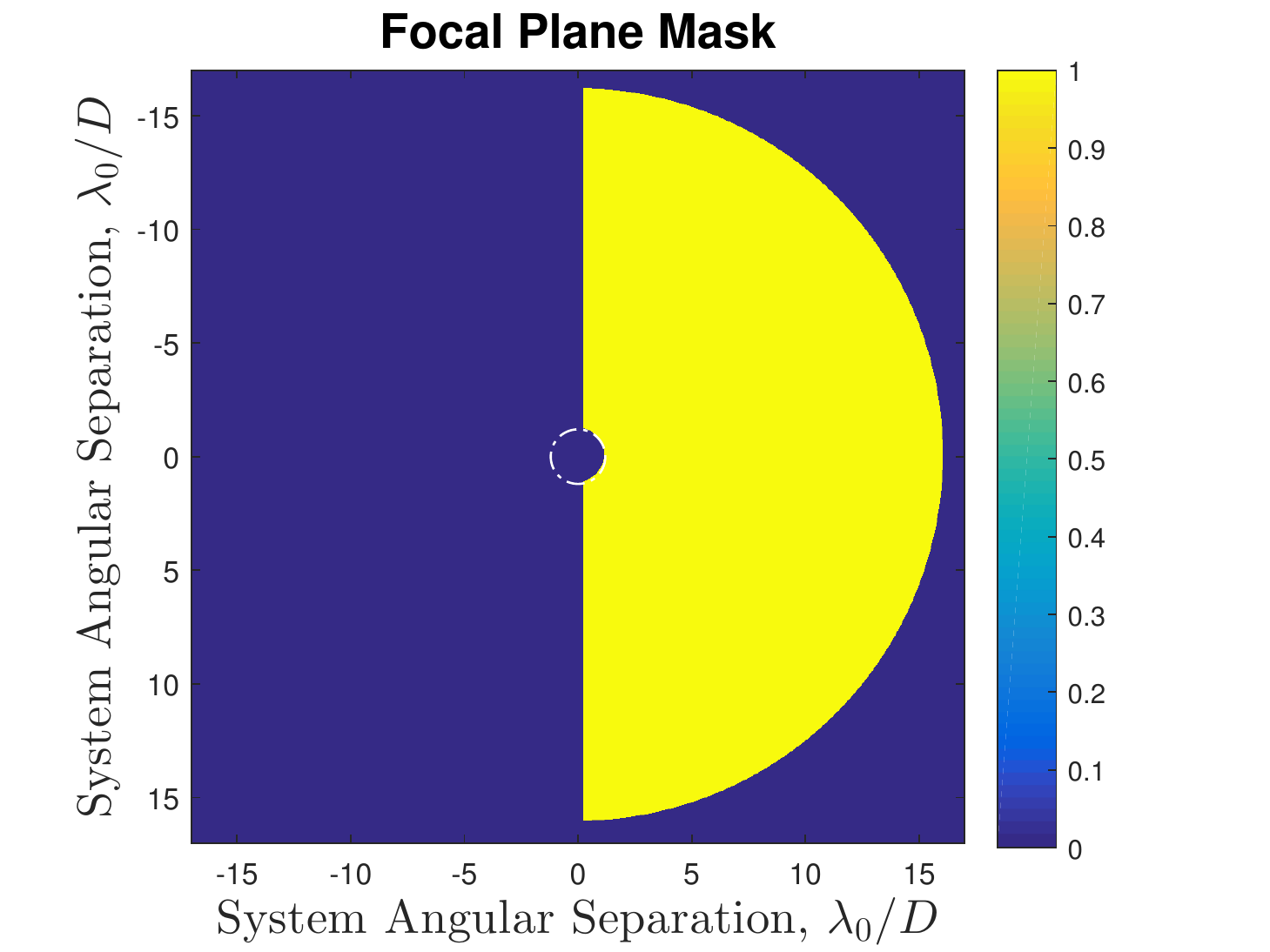}
}
\caption[Summary of the EXCEDE mission with telescope and science capabilities.]{\label{fig:piaa} \subref{fig:piaa-m2} The simulated PIAA pupil (M2 plane) \subref{fig:piaa-fpm} The one-sided C-shaped focal plane mask.}
\end{figure}

The EXCEDE layout contains forward PIAA coronagraphic optics with two mirrors. These mirrors are defined by two planes at M1 and M2. The propagation between M1 and M2 is defined in our model by a ray-tracing pupil-mapping function. The intensity at the forward PIAA exit pupil (M2) is shown in Figure \ref{fig:piaa}\subref{fig:piaa-m2}. The DM is physically located upstream of the PIAA system and in the simulation this is approximated as the DM being conjugate to M1. 

The C-shaped Focal Plane Mask (FPM) as shown in Figure 13 (right), is located at the first focal plane F1. The propagation between the exit PIAA pupil at M2 and the entrance pupil of the first inverse PIAA lens at L2 is performed using a convolution operation (with a Fourier transformation of the FPM). The inverse PIAA lenses at L2 and L1 perform an inverse pupil-remapping operation to the forward PIAA mirrors also modeled through ray-tracing. At the exit pupil of the inverse PIAA system, a Lyot stop blocks diffracted light from the focal plane mask. The physical dimension of the PIAA system, the corresponding magnification and sizing of the focal plane mask, the open diameter of the Lyot stop, and the pixel sampling at the final science plane are all matched to the experimental testbed.

All optical aberrations are collocated at the entrance pupil of the system, the M1 plane, and are propagated through the system as described. These aberrations can be corrected with the DM using either the iterative Electric Field Conjugation (EFC) or Speckle Nulling (SN) wavefront control algorithms. As in the VCT 5 demonstration, in simulation the WFC are applied for the central wavelength of 650 nm, the DM setting is maintained, and the input light bandwidth is extended to $\lambda_\textrm{central} / \Delta \lambda$ = 10\%. This procedure follows the experimental correction. 

The geometrical remapping method has been validated against other PIAA propagators (geometrical remapping with Talbot effect correction which simulates diffraction, Fresnel diffraction, S-Huygens diffraction \cite{Belikov06}) with the following conclusions: for on-axis modes with no errors, geometrical remapping gives the same result (down to $10^{-10}$ level) as long as the edges of the pupil are feathered with a pre (or post-) apodizer. Without apodizers, the results are correct to the $10^{-7}$ contrast level of this experimental demonstration. For low-order modes (tip/tilt, defocus, etc) the results are similar. For higher order modes, the shape of the speckle field in the focal plane starts deviating between geometrical remapping and other models, but the contrast levels remain the same. Therefore, for purposes of determining contrast of the speckles (as opposed to exact morphology), geometrical remapping is a sufficient model for the contrast levels at which the first-generation PIAA system used in this experiment operates.

\subsection{Ideal Performance} \label{sect:ideal}

\begin{figure}
\centering
\includegraphics[height = 3.5 in]{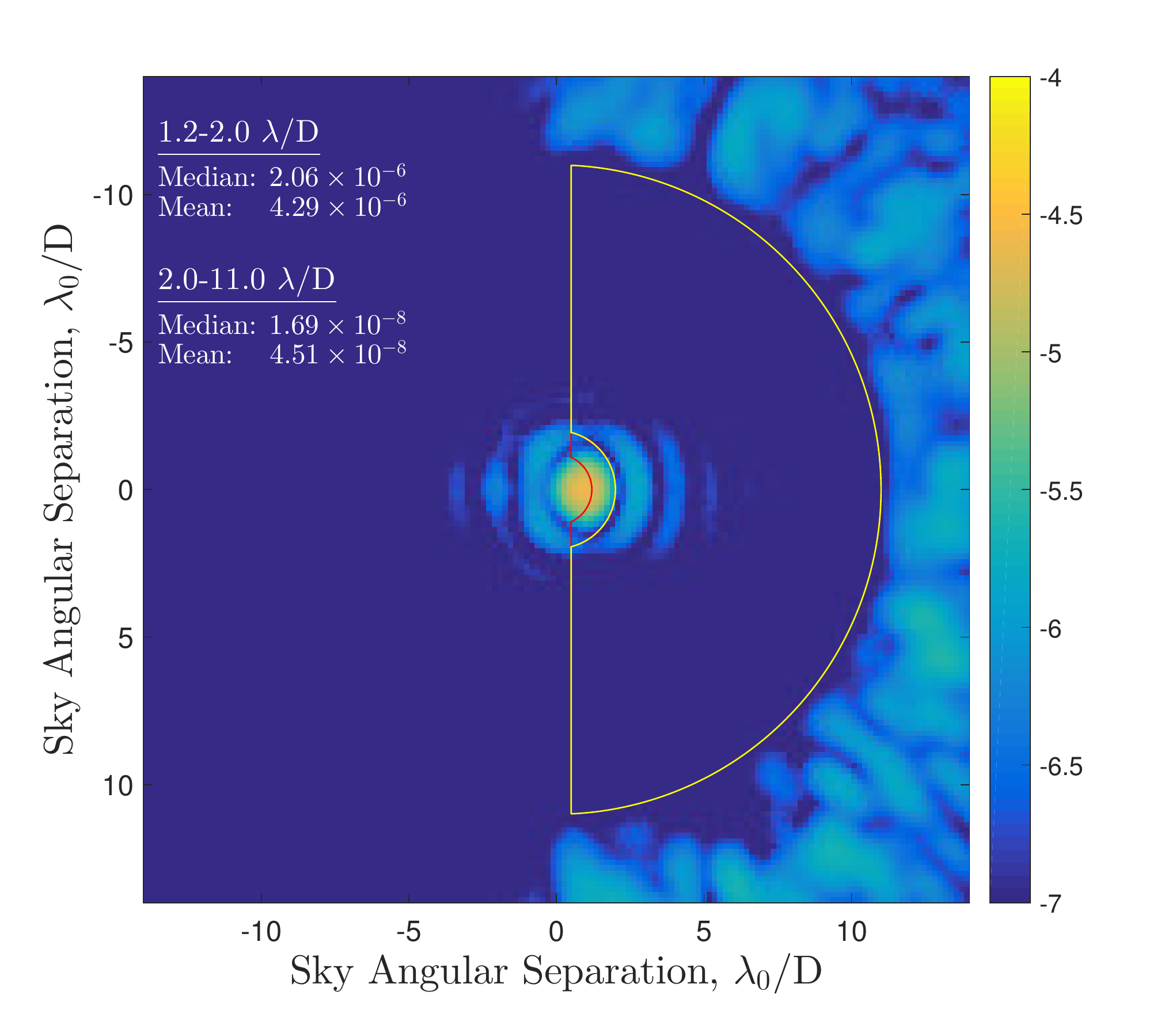}
\caption[Ideal testbed performance.]{\label{fig:ideal} Ideal performance of EXCEDE starlight suppression system in simulation with closed-loop wavefront correction. The theoretical best performance in the IWZ and OWZ for the EXCEDE laboratory bench is computed in terms of both median and mean raw contrast.}
\end{figure}

To establish a baseline against which to compare the observed limitations of the experiment, we simulate the performance of the system first under ideal conditions. We first simulate propagation through the system and application of wavefront control without any aberrations. This involves generating only a pure $\lambda$/20 RMS phase aberration with a decreasing frequency ramp ($1/f^{3/2}$ mimicking typical optical shape errors at high frequencies \cite{Krist08}) in amplitude and assuming the circular occulter of the focal plane mask is completely opaque. In this case, the starting contrast in the dark hole region is limited only by speckles caused by relatively high frequency scattering giving rise to a PSF with a near-unity Strehl Ratio of 0.99. The result of this ideal case for 10\% broadband light  is shown in Figure \ref{fig:ideal} after wavefront control is applied. Thus, the median contrast in 10\% light is $2.06 \times 10^{-6}$ from 1.2-2.0 $\lambda$/D (IWZ) and $1.69 \times 10^{-8}$ from 2.0-11 $\lambda$/D (OWZ). This represents, within two significant figures, recovery of an ideal (completely aberration-free) system; thus $\lambda$/20 errors are completely correctable to these levels of contrast in our testbed. We have compared both Speckle-Nulling (SN) and Electric-Field Conjugation (EFC) wavefront correction to ensure results are not algorithm-limited and obtained near-identical final performance. Introduction of experimental limitations will worsen contrast performance compared to this ideal level. 

\subsection{Low Order Aberrations}

The EXCEDE starlight suppression system has a LOWFS to measure low-order aberrations with commands sent to the DM for their correction. In this technology demonstration, the low-order aberrations that are sensed and corrected are the tip/tilt modes. In ex post-facto model simulations, we introduce phase aberration modes in proportions that match the shape and Strehl Ratio (SR) of the experimental PSF. This is achieved by applying the Gerchberg-Saxton algorithm \cite{GerchbergSaxton} to compute an estimate of the net phase aberrations through the entire system.

\begin{figure}[t!]
\subfloat[]{
\includegraphics[height = 1.5 in]{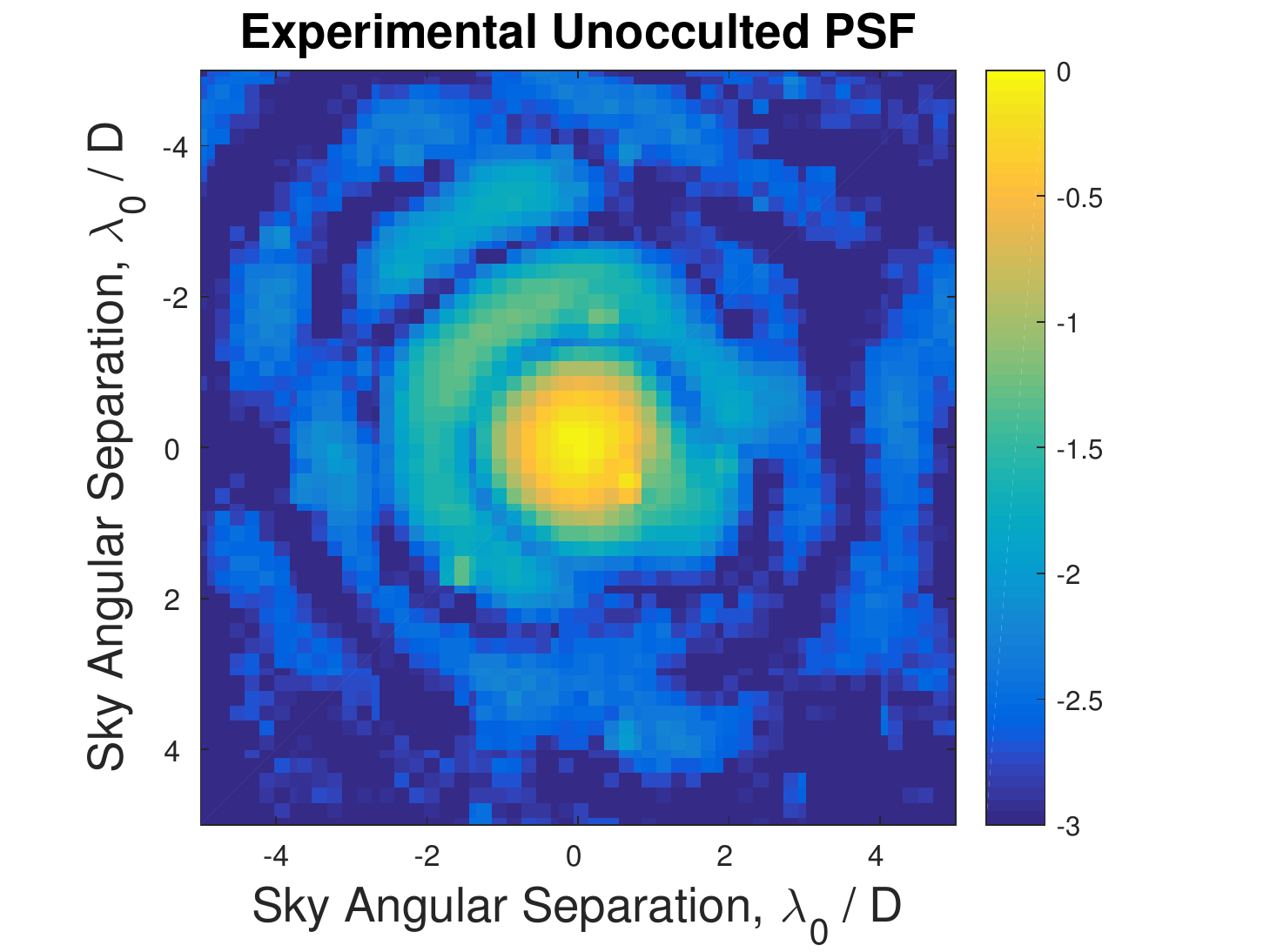}
\label{fig:psf-exp}
}
\subfloat[]{
\includegraphics[height = 1.5 in]{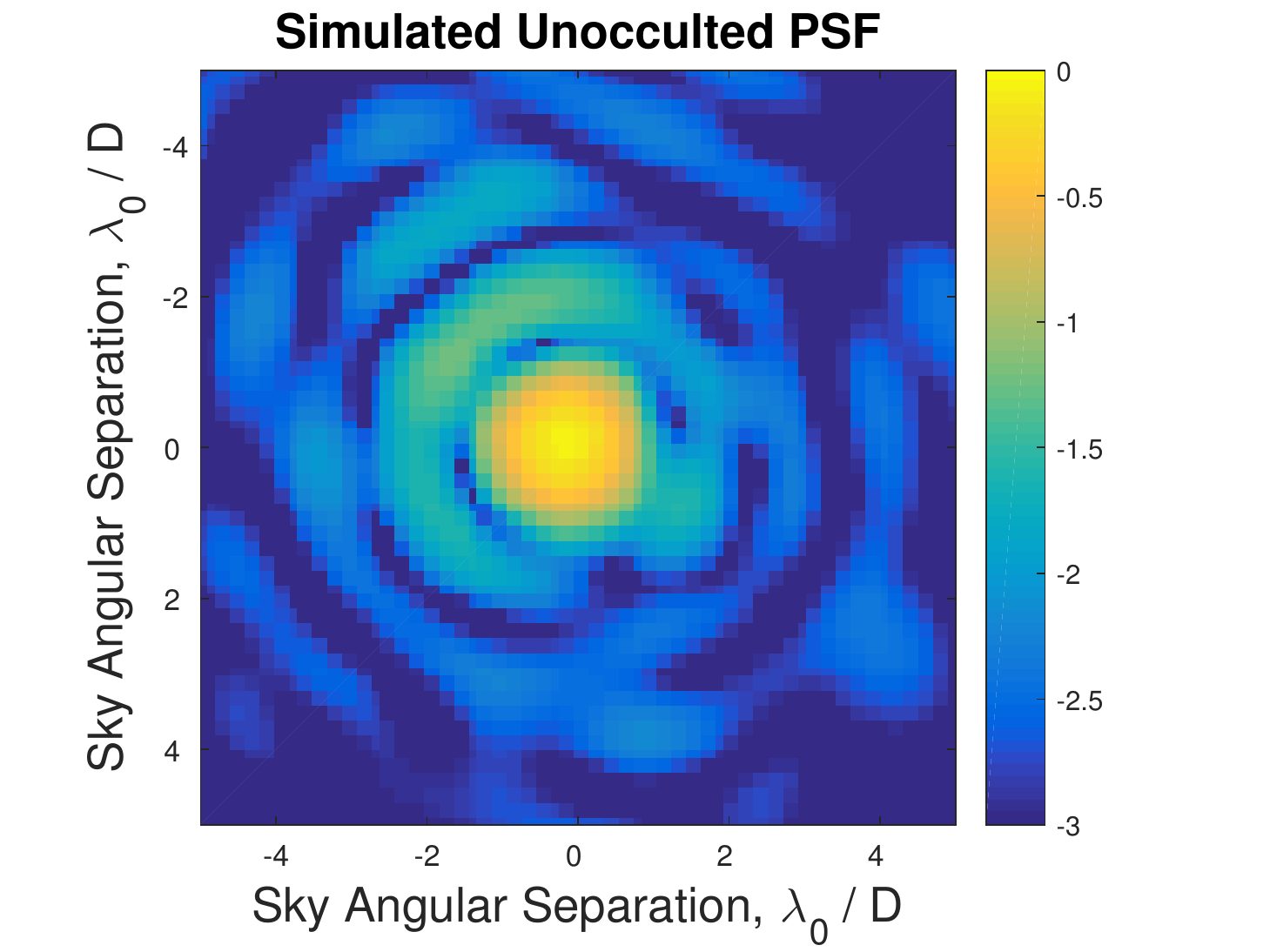} 
\label{fig:psf-sim}
}
\subfloat[]{
\centering
\includegraphics[height = 1.5 in]{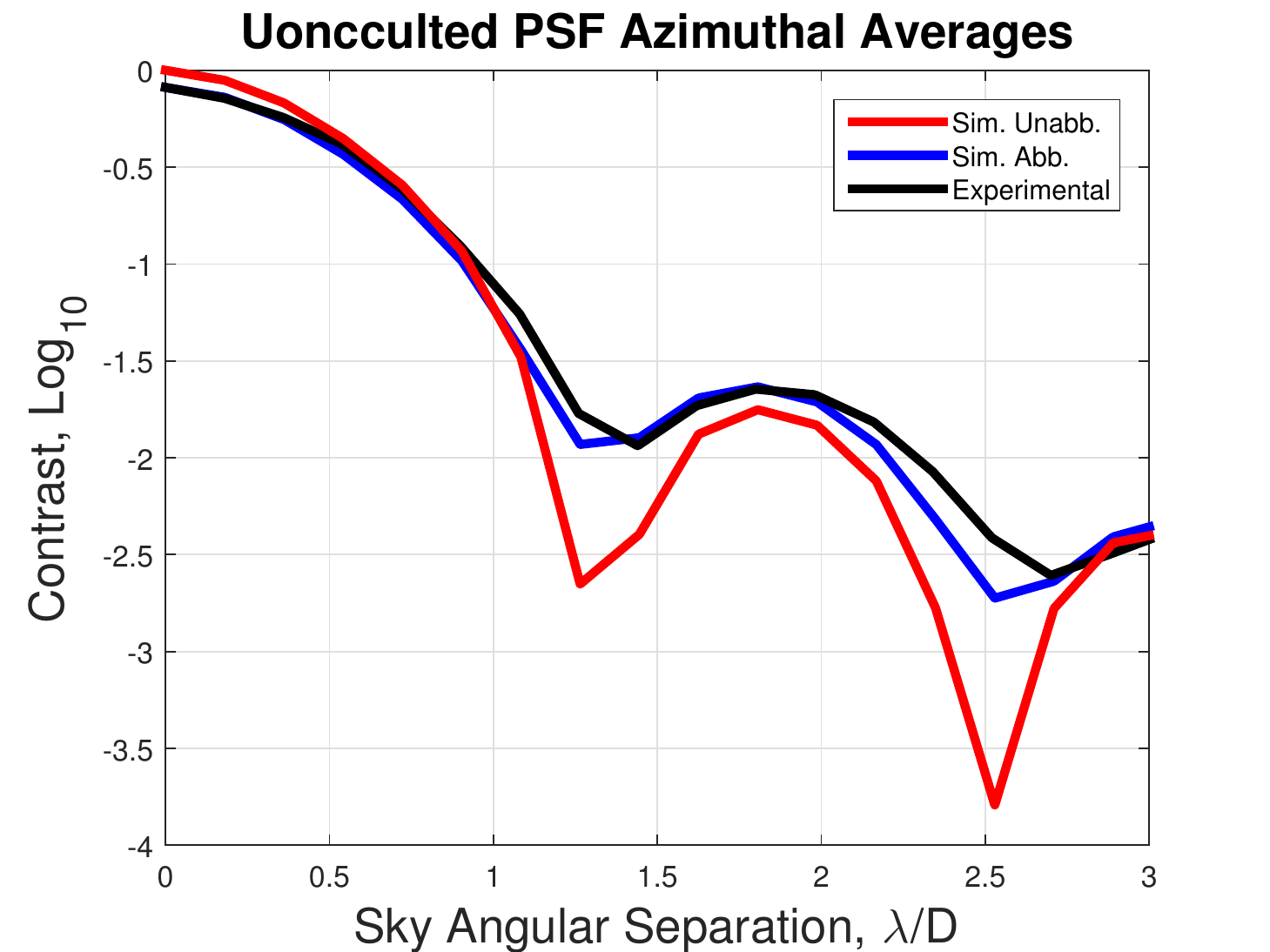}
\label{fig:psf-az}
}
\caption[Comparison of the experimental and simulated PSFs.]{\label{fig:psf} \subref{fig:psf-exp} Experimentally measured on-axis PSF without an occulter (SR = 0.81); the dot artifacts are contaminant deposition on the inside of the CCD window (see description in \S \ref{sec:testbedDescription}) \subref{fig:psf-sim} Simulated on-axis PSF matched to the experimental PSF \subref{fig:psf-az} Azimuthally averaged profiles of the unocculted PSFs comparing the unabberated simulated system (red and blue curves respectively) and the best-matched aberrated simulated system with the experimental results (black curve).}
\end{figure}

In Figure \ref{fig:psf}, we compare images of the experimental PSF in \subref{fig:psf-exp} with the corresponding simulated PSF \subref{fig:psf-sim} -- both represent the on-axis, unocculted case. The simulated PSF corresponding to the estimated aberrations phase-map corresponds to a SR of 0.81 which matches experimental Strehl measurements taken during alignment (0.8-0.85). We also show in the right-pane 360-degree azimuthally averaged radial profiles of the experimental and simulated PSFs. The red-curve represents the unaberrated theoretical (ideal) PSF of the system. There is very good agreement, especially in regions corresponding to a low spatial-frequencies, in the PSF between the simulated aberrated PSF and the experimental PSF as seen by azimuthal comparison of the corresponding blue and black curves.

\begin{figure}
\centering
\includegraphics[height = 2.5 in]{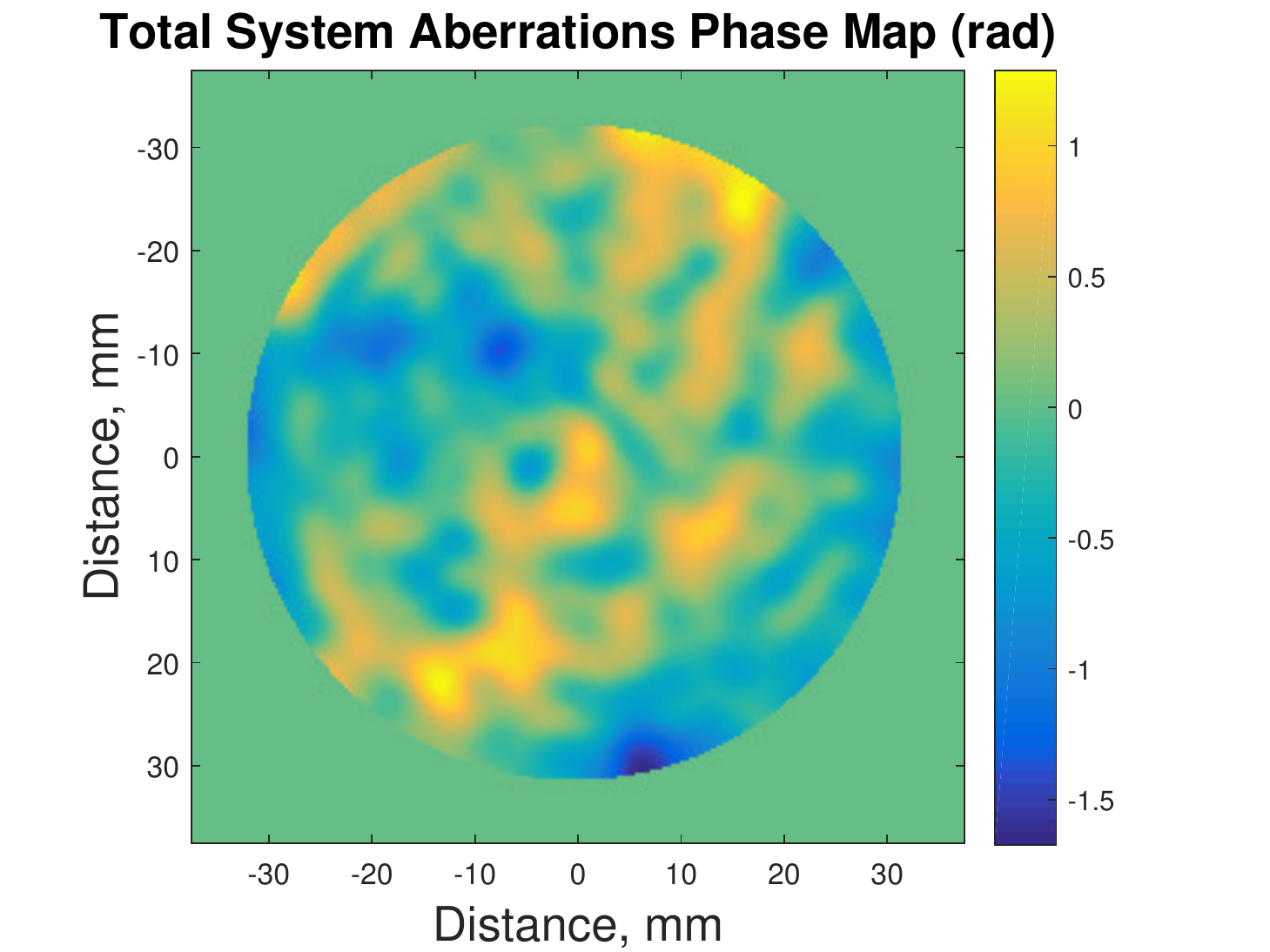}
\caption[Total phase aberrations present in the optical system. ]{\label{fig:abPup} Phase aberrations obtained from applying the Gerchberg-Saxton algorithm on the unocculted PSF and distributed between the entrance PIAA pupil (M1) and the Inverse PIAA exit pupil (L1) in the simulation. Phase aberration RMS value is  0.4 rad. }
\end{figure}

This representative simulated PSF is obtained by applying the Gerchberg-Saxton algorithm on an unocculted PSF image used to calibrate contrast in the experimental data set for 300 iterations until the phase estimate converges. The resulting total phase aberration map across the entire optical system is shown in Figure \ref{fig:abPup}. These aberrations correspond to RMS amplitude of 0.4 radians. However, measurements of the aberrations are not available at individual optical planes. Therefore this total phase map is distributed roughly equally upstream of the focal plane mask (applied at the PIAA entrance pupil  -- M1) and downstream of the focal plane mask (applied at the Inverse PIAA exit pupil -- L1). Fitting Zernike polynomials to these phase aberrations shows that they are dominated by the first 30 Zernike polynomials, with the presence of primary and secondary astigmatism resulting in a good match of the diffracted light beyond the focal plane mask. The cause of these astigmatic aberrations remains somewhat speculative. We suggest that the most likely source is due to misalignment in the front-end OAPs. The optical alignment of the testbed was performed in air, then tested in vacuum, and misalignment (particularly between the OAPs) plausibly was introduced by the change of environment from air to vacuum.

We then perform a sensitivity analysis with respect to a residual tip/tilt term. No tip/tilt terms were added with the low-order aberrations, but after performing wavefront control and obtaining an optimal DM setting we add residual tip/tilt to determine the robustness of the solution to small tip/tilt deviations. Because of the number of iterations involved in a typical experimental run and the integration times, it is likely that some residual misalignment may develop. 

\begin{figure}[t!]
\centering
\subfloat[]{
\centering
\label{fig:sensitivity-tilt}
\includegraphics[height = 2.45 in]{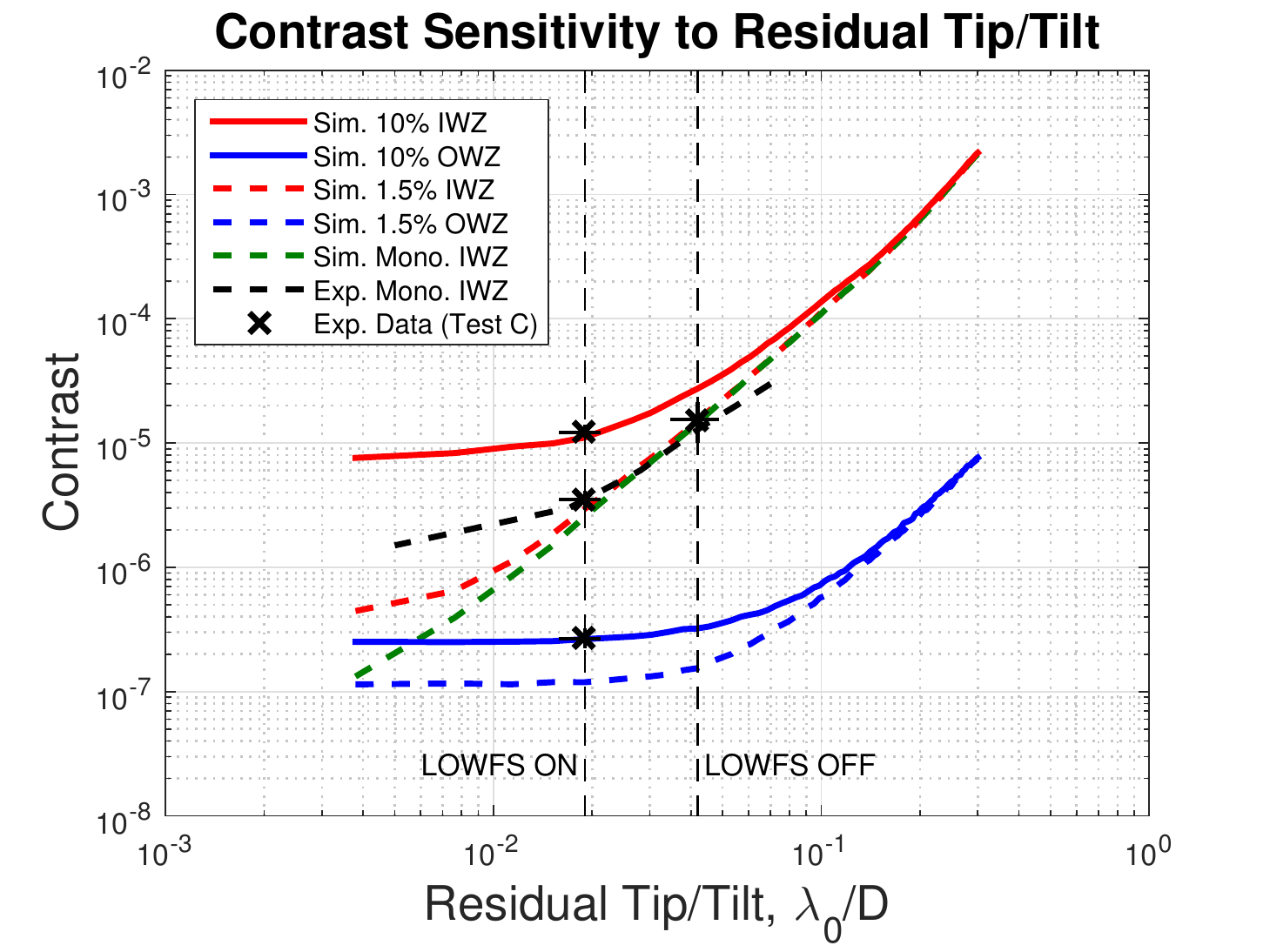}
}
\subfloat[]{
\centering
\label{fig:sensitivity-bandwidth}
\includegraphics[height = 2.45 in]{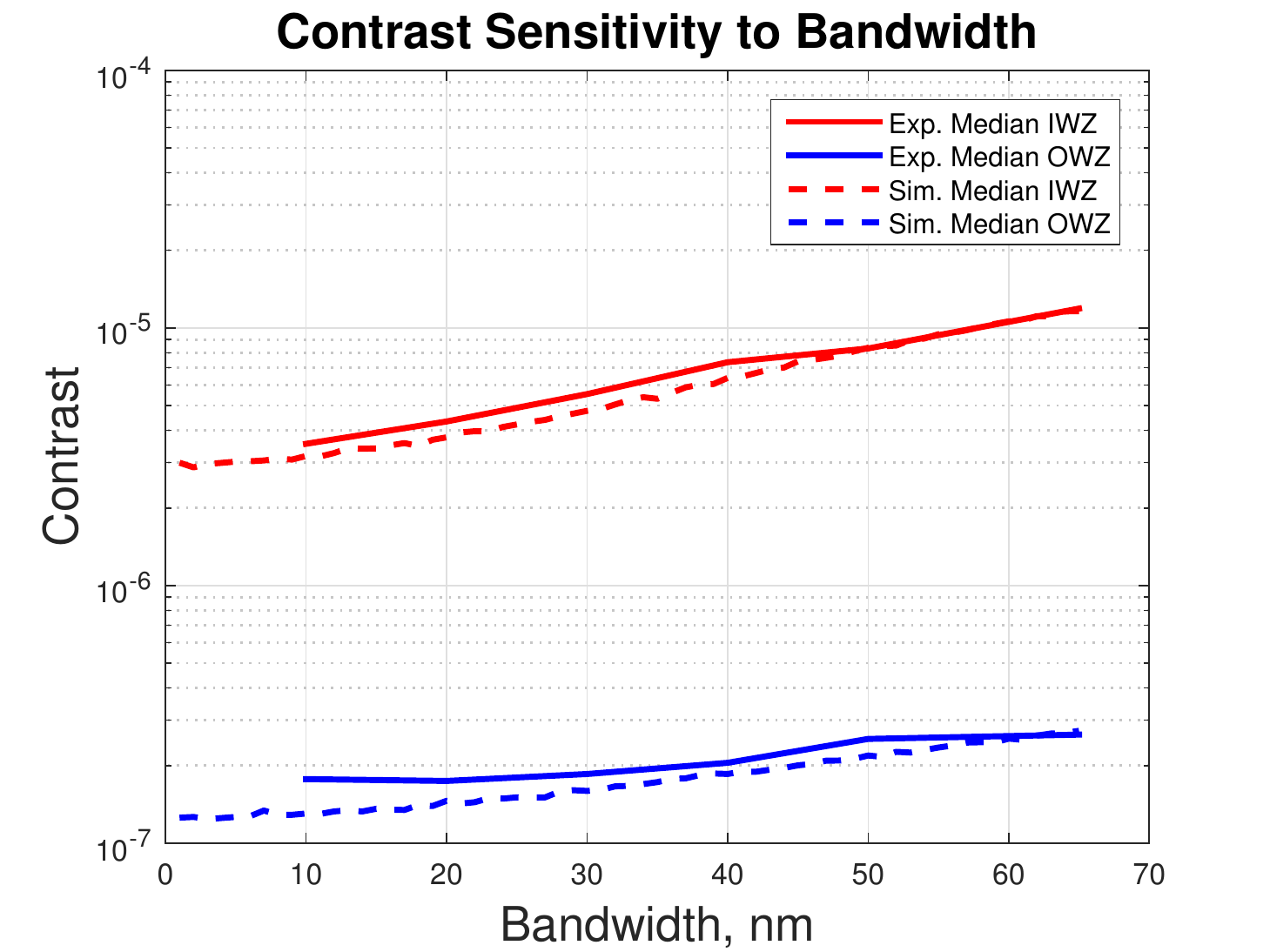}
}
\caption[Contrast sensitivity as a function of residual tip/tilt and bandwidth.]{\label{fig:sensitivity} Contrast sensitivity analysis of model for \subref{fig:sensitivity-tilt} residual tip/tilt showing the simulated model at 1.5\% (dashed lines) and 10\% (solid lines) for both IWZ (red lines) and OWZ (blue lines) with experimental data points collected during Test A indicating performance prior to engaging the LOWFS and after the contrast level is stabilized with the LOWFS turned on; additionally a separate experimental data set was collected using monochromatic input light (dashed black line) and this is compared to simulated monochromatic contrast performance (dashed green line); \subref{fig:sensitivity-bandwidth} comparison of contrast sensitivity as a function of spectral bandwidth in simulation with the experimental data collected in Test C.}
\end{figure}

The results of the sensitivity to residual tip/tilt are shown in Figure \ref{fig:sensitivity}\subref{fig:sensitivity-tilt}. We plot both the 1.5\% bandwidth and the 10\% bandwidth simulated sensitivity contrast curves for both the IWZ and OWZ. We compare the simulated contrast IWZ contrast curve to an experimental sensitivity test (black dashed curve) for which we injected artificial tip/tilt on the DM and the resulting contrast was measured. Although there is a discrepancy in the exact contrast level the two contrast decades per decade of residual tip/tilt is matched between between the experiment and the simulated model. Finally, we indicate the contrast levels obtained from the Test A experimental run. In the experimental procedure described in \S \ref{sect:expResults}, the SN wavefront correction algorithm was run for monochromatic input light until a $1 \times 10^{-5}$ contrast level was reached in the IWZ. Then the LOWFS was turned on and the SN WFC loop operated in tandem with the LOWFS. Subsequently after allowing the SN WFC to operate with LOWFS, the monochromatic contrast level in the IWZ improved to $3 \times 10^{-6}$. Once this level stabilized the DM settings were maintained constant, the input light bandwidth was increased to 10\%, and the corresponding broadband contrast was measured in both the IWZ and OWZ. As indicated in Figure \ref{fig:sensitivity}\subref{fig:sensitivity-tilt}, these experimental contrast data points lie on the simulated sensitivity curves and give a measure of the improvement from the removal of the tip/tilt modes by the LOWFS.

A question we investigate is whether the observed degradation in contrast performance as a function of spectral bandwidth (as shown in the experimental bandwidth sensitivity at the end of Test C in Figure \ref{table:experimentalResults}) is captured by the model. We computed the IWZ and OWZ median contrast by maintaining the obtained narrowband DM settings gradually increasing the filter bandwidth from 10 nm (1.5\%), the minimum tunable setting in our test configuration, up to 65 nm (corresponding to 10\% light) or vice-versa. This was also done in simulation. The comparison is shown in Figure \ref{fig:sensitivity}\subref{fig:sensitivity-bandwidth}. For the IWZ, we have a factor of 3 degradation of contrast from 10 nm to 65 nm in the test bed that is matched by simulation. The degradation in contrast for increased bandwidths is more pronounced for the IWZ than for the OWZ.

\subsection{Simulation Results}

\begin{figure}[t!]
\centering
\subfloat[]{
\centering
\label{fig:compareResults-sim}
\includegraphics[height = 2.75 in]{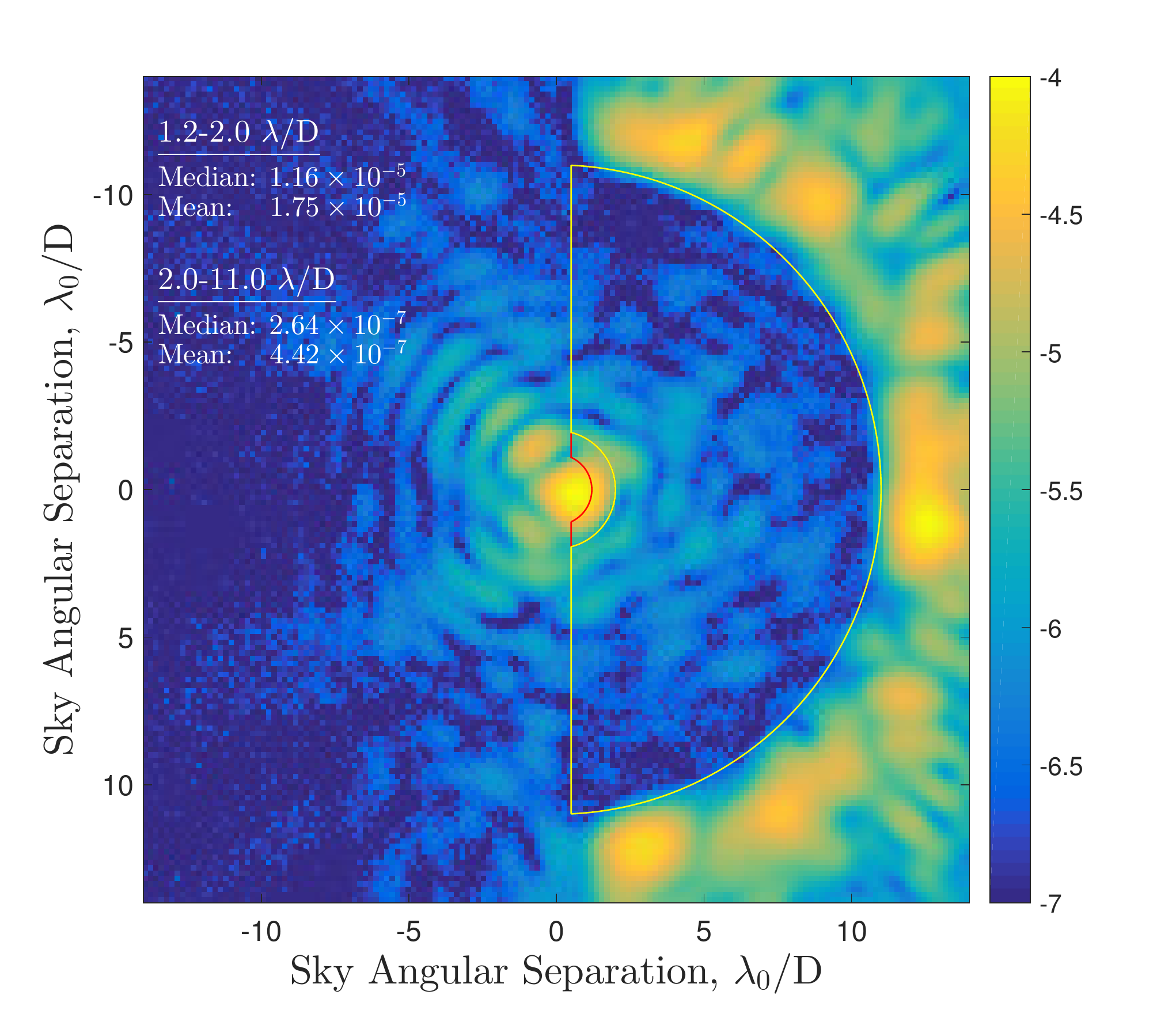}
}
\subfloat[]{
\centering
\label{fig:compareResults-exp}
\includegraphics[height = 2.75 in]{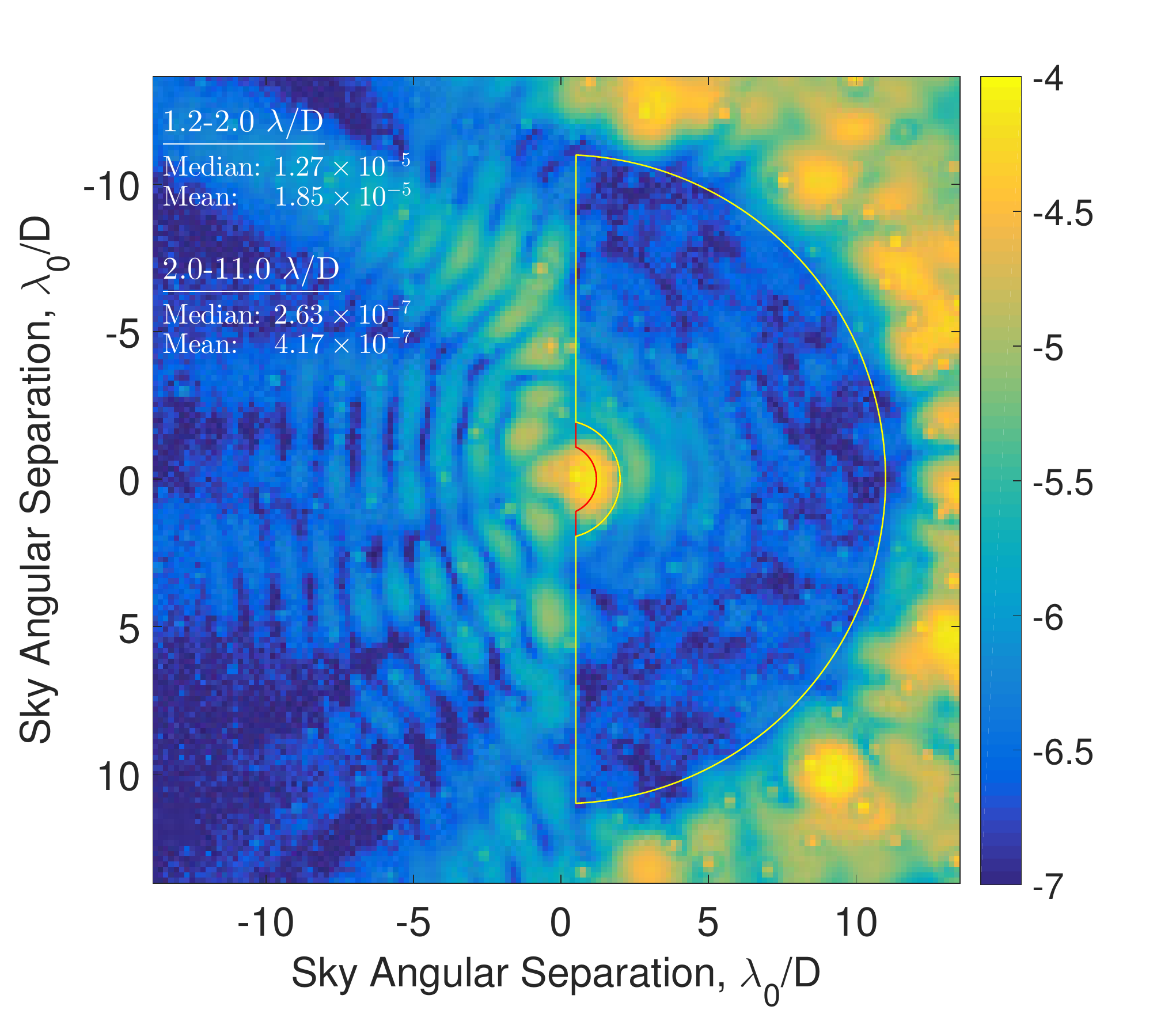}
}
\caption[Comparison of simulation and experimental results.]{\label{fig:compareResults} Performance comparison of the EXCEDE optical testbench between: \subref{fig:compareResults-sim} simulation results \subref{fig:compareResults-exp} experimental results.}
\end{figure}

To determine the effect of the low-order aberrations on the performance of the EXCEDE starlight suppression system, we perform a simulation similar to the ideal case that established a performance baseline. In particular, we apply the low-order aberrations matched to the measured experimental PSF as described in the previous section together with $\lambda$/20 surface aberrations as expected from the optical surfaces. The result of this simulation is shown in Figure \ref{fig:compareResults}\subref{fig:compareResults-sim}. We compare this result with experimental results, in particular a representative frame from Test A in Figure \ref{fig:compareResults}\subref{fig:compareResults-exp}. The contrast in the IWZ is limited by diffracted light at the center of the image. The shape and intensity of this diffracted light is related to the low-order aberrations present in the system upstream of the focal plane occulter (at the PIAA M1 entrance pupil plane). We can observe important qualitative similarities between the simulation and experiment. Specifically, the diffracted light centered behind in the focal plane occulter features a bright spot in the center and two dimmer companion spots. The location of the spot is matched to that of the experiment by the inner working angle verification routine -- when the mask is set such that the IWA is located at 1.2 $\lambda$/D, the centroid of the simulated diffraction spot is the same as that observed in the experiment. The ringing structure around this bright spot is at a similar scale (approximately 4 Airy rings out to 5 $\lambda$/D), and this is given by matching the size of the Lyot stop between experiment and simulation. The contrast field in the OWZ contrast level is well-matched between simulation and experiment although there are some morphological differences in the speckle structure.

\begin{figure}[t!]
\centering
\includegraphics[height = 3.5 in]{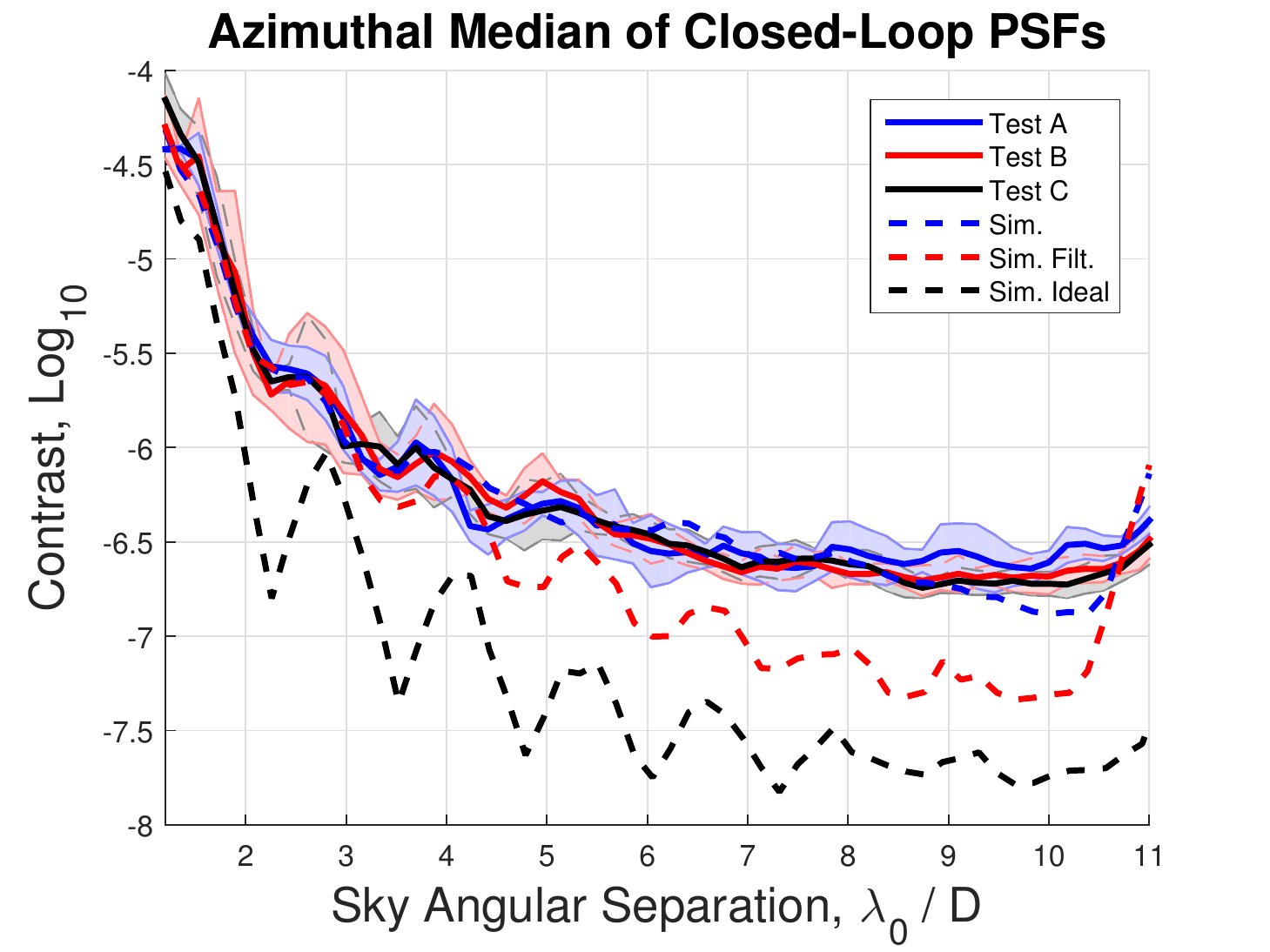}
\caption[Azimuthal comparison of the experimental and simulated PSFs azimuthal.]{\label{fig:azimuthalCompare} Azimuthal comparison of the median for the three experimental tests indicated by solid curves and simulation. Shaded regions represent $\pm$15th percentile about the median for the experimental results and show contrast spread. Three simulation curves are depicted by dashed lines: the simulated ideal case represents the best-case performance attainable with this configuration, the simulated filtered case uses aberrations representing only the first 30 Zernike terms which dominate performance in the IWZ, and the combined simulation with the addition of higher mid-spatial frequency provides a good match to experimental results. }
\end{figure}

For a more quantitative comparison, we plot azimuthal median curves in Figure \ref{fig:azimuthalCompare}. We show the azimuthal median of each of the three reduced experimental tests as solid colored curves, with the 30th percentile population around the median indicated in the shaded region as an indicator of the contrast spread. This demonstrates the repeatability of the experiment not only in terms of the overall contrast level but also in terms of spatial distribution across the entire dark hole. The azimuthal median of the simulated system is shown via the dashed blue curve and is in good agreement with the experimental data. As an illustration of the deterioration of contrast due to the presence of low-order aberrations we have also plotted the azimuthal curve of the ideal system simulation previously shown in Figure \ref{fig:ideal}.

Filtering of the phase aberrations to keep only the first 30-fitted Zernike polynomial terms (dashed red line in Figure \ref{fig:azimuthalCompare}) maintains good agreement in the IWZ but the overall simulated contrast in the OWZ is nearly an order of magnitude better than the level observed experimentally. This indicates that, as expected, the IWZ contrast is primarily limited by the low-order phase aberration modes with the OWZ limited by mid-spatial frequency modes. Improvements in optical alignment in vacuum to reduce low-order aberrations (for example, using a criterion that the starting SR should be 0.9 or better) and through estimation and correction of additional low-order modes with the LOWFS is expected to improve the IWZ contrast to approach the EXCEDE science goal of $10^{-6}$ raw median contrast in the IWZ.

It is worth mentioning here that we only considered pure phase aberrations insofar in this study, but for other experiments amplitude aberrations from the BMC DM were found to be a limiting factor \cite{Mazoyer14}. To determine whether amplitude aberrations can become limiting factors, we introduced amplitude aberrations instead of pure phase aberrations in the closed-loop simulation. We found that for amplitude aberrations with RMS reflectivities less than 0.05 (and using a $f^{-3/2}$ power law), the ideal contrast performance levels in \S \ref{sect:ideal} were completely recoverable in the one-sided DH used in this study with WFC in simulation. For amplitude aberrations to be a limiting factor it was necessary to introduce amplitude aberrations with RMS reflectivities at the RMS level of 0.15 or greater. Additionally, these amplitude aberrations were limiting at the deepest contrast level in the OWZ rather than in the IWZ where we found our performance to be limited. Although operating with a different coronagraph, this conclusion is consistent with the earlier study \cite{Mazoyer14} which was operating at a deeper contrast level and greater inner working angles. Finally, the BMC DM was found to be capable of reaching deeper contrast levels with a PIAA coronagraph than in this study at the $10^{-8}$ level reported in-air at Ames (but a larger inner working angle) \cite{Belikov12} -- again indicative of the fact that amplitude aberrations are not limiting at the contrast levels and working angles in our study.

\section{Conclusions}
\label{sect:conc}

In this paper, we have presented experimental results from the EXCEDE starlight suppression system demonstrating wavefront control high-contrast imaging capability in 10\% broadband light centered about 650 \si{\nano \meter}. This experiment was conducted in a space-relevant vacuum environment for small angular separations from 1.2 to 11 $\lambda$/D using a highly-efficient PIAA coronagraph architecture closely approximating the conditions for the proposed EXCEDE mission. We have measured through three fully independent experimental runs repeatable and stable median raw per pixel median image contrast of $1 \times 10^{-5}$ from 1.2-2.0 $\lambda$/D simultaneously with $3 \times 10^{-7}$ from 2.0-11.0 $\lambda$/D. These experimental results are close to the EXCEDE requirements. 

Numerical simulations of our optical model of the EXCEDE system show a good match to experimental results for dependence on spectral bandwidth and residual tip/tilt modes. We have investigated the degradation in contrast from the ideal performance case showing that the contrast in the inner working zone is limited to the observed experimental levels by low-order aberrations including tip/tilt residuals and their combination with optical errors. These simulation results represent a validation of the optical models used to design the EXCEDE SSS, predict its performance, and understand the laboratory limits . 

The work has described results that are part of technology maturation process for the EXCEDE coronagraph, however it also has significance relevant to other missions. In particular the WFIRST-AFTA PIAACMC architecture has elements in common with some of those explored with the EXCEDE testbed. Additionally, through experimental verification for EXCEDE, our results explore fundamental trade-offs between contrast, inner working angle, and sensitivity to low-order aberrations and thus this work is relevant to all Lyot-style coronagraphs helping to raise the technology readiness level for high-contrast starlight suppression.


\acknowledgments 
This work has been supported, in part by a NASA Explorer category 3 program grant \# NNX12AH39G  to the University of Arizona,  NASA's Ames Research Center, and the Technology Development for Exoplanet Missions (TDEM) program through solicitation NNH09ZDA001N-TDEM at NASA's Science Mission Directorate. The material is based upon work supported by NASA under Prime Contract Number NAS2-03144 awarded to the University of California, Santa Cruz, University Affiliated Research Center. It was carried out at the NASA Ames Research Center and the Lockheed Martin Advanced Technology Center. 


\bibliography{database}   
\bibliographystyle{spiejour}   


\vspace{2ex}\noindent\textbf{Dan Sirbu} is an NPP postdoctoral researcher at the NASA Ames Research Center and is affiliated with the Ames Coronagraph Experiment (ACE) Lab. He received his BSc in Electrical Engineering from the University of Alberta in 2008, and his PhD degree in Mechanical \& Aerospace Engineering from Princeton University in 2014. His research interests are related to high-contrast imaging for the detection and characterization of exoplanets including advanced wavefront control and estimation algorithms, internal coronagraphs, and external occulters. 



\vspace{1ex}
\noindent Biographies and photographs of the other authors are not available.

\listoffigures
\listoftables

\end{spacing}
\end{document}